\documentclass[11pt]{article}
\usepackage{amsmath,amssymb,epsfig,sint}

\newcommand{\R}{\rm I\kern-.2emR}
\newcommand{\C}{\rm \kern.25em\vrule height1.4ex
depth-.12ex width.06em\kern-.31em C}
\newcommand{\N}{{\rm I\kern-.16em N}}
\newcommand{\Z}{{\rm Z\kern-.35em Z}}

\newcommand{\rme}{{\rm e}}
\newcommand{\rmd}{{\rm d}}
\newcommand{\rmO}{{\rm O}}

\newcommand{\ch}{{\rm ch}}
\newcommand{\sh}{{\rm sh}}
\newcommand{\fm}{{\rm fm}}

\newcommand{\be}{\begin{equation}}   
\newcommand{\ex}{\end{equation}}
\newcommand{\ba}{\begin{eqnarray}}
\newcommand{\ea}{\end{eqnarray}}

\newcounter{subequation}[equation]
\makeatletter

\expandafter\let\expandafter
\reset@font\csname reset@font\endcsname

\def\subeqnarray{\arraycolsep1pt
    \def\@eqnnum\stepcounter##1{\stepcounter{subequation}%
        {\reset@font\rm(\theequation\alph{subequation})}}
\jot5mm     \eqnarray}

\makeatother

\newcommand{\msbar}{{\rm \overline{MS\kern-0.14em}\kern0.14em}}

\begin{document}
\begin{titlepage}

\begin{flushright}
   MPP-2008-38\\
   UTHEP-562\\
   May 2008
\end{flushright}

\vskip 1.0 true cm
\begin{center}
{\Large\bf 
Bethe--Salpeter wave functions in integrable models}
\\
\vspace{1cm}
\end{center}

\vskip 1.0 true cm
\centerline{\large Sinya Aoki}
\vskip1ex
\centerline{Graduate School of Pure and Applied Sciences, 
University of Tsukuba}
\centerline{Ibaraki, 305-8571, Japan}
\centerline{Riken BNL Research Center, Upton, NY 11973, USA}
\vskip 0.7 true cm
\centerline{\large Janos Balog}
\vskip1ex
\centerline{Research Institute for Particle and Nuclear Physics}
\centerline{1525 Budapest 114, Pf. 49, Hungary}
\vskip 0.7 true cm
\centerline{\large Peter Weisz}
\vskip1ex
\centerline{Max-Planck-Institut f\"ur Physik}
\centerline{F\"ohringer Ring 6, D-80805 M\"unchen, Germany}
\vskip 1.0 true cm
\centerline{\bf Abstract}
\vskip 1.0ex
We investigate some properties of Bethe--Salpeter wave functions 
in integrable models. In particular we illustrate the application 
of the operator product expansion in determining the short distance 
behavior. The energy dependence of the potentials obtained from such 
wave functions is studied, and further we discuss the (limited) 
phenomenological significance of zero--energy potentials.

\vfill
\eject

\end{titlepage}

\section{Introduction}

In a recent paper \cite{IAH1} which has received general recognition
\cite{Nature2007}, Ishii, Hatsuda and one of the present authors (S.~A)
have presented results on the nucleon--nucleon (NN) potential from 
first principle lattice computations \cite{IAH2,SinyaLATT07}.
The results qualitatively resemble phenomenological NN potentials
which are employed in nuclear physics. The force at
medium to long range ($r\ge 1.2\fm$) is attractive; this feature which is 
essential for the existence of bound states of nuclei (e.g. the deuteron) 
has long well been understood in terms of pion and other heavier meson 
exchange. At short distances a characteristic repulsive core is produced 
\cite{IAH1} by the QCD dynamics, but this feature has not yet found
a simpler theoretical explanation.  

Intuitively the short distance behavior in QCD is encoded in operator
product expansions (OPE). However wave functions and potentials in the 
framework of relativistic quantum field theory are notoriously 
``flexible" concepts. There are infinitely many definitions depending 
on the interpolating fields chosen and thus the universality of the 
short distance behavior extracted from one particular chosen wave function 
is not a priori clear.
The wave function discussed in ref.~\cite{IAH1} is a Bethe--Salpeter 
(BS) wave function with a particular nucleon interpolating field of lowest 
dimension. The phenomenological success of the results gives rise to
the hope that one is on the ``correct track", however there remain many
theoretical questions and refinements in the measurements to be made.
For example the results are still in the quenched approximation,
lattice artifacts must be studied in more detail, 
the dependence of the results on the interpolating field must be 
examined and the very definition of a potential via a BS wave function 
must be better understood.

It is our hope that studies of BS wave functions in integrable models
in two dimensions will give us more insight into such questions.
As an aside here we note that the methods used in ref.~\cite{IAH1}
were partially motivated by a method proposed to measure phase shifts 
in a two--dimensional model \cite{BNNPSW}. In this paper we investigate
BS wave functions in the Ising model and in the O(3) non--linear 
sigma--model in two space--time dimensions. 

In a remarkable paper, Fonseca and Zamolodchikov \cite{FZ}
obtained an exact expression for the BS wave function of the Ising 
field theory. In Sect.~2 we study its properties; 
in particular we can see at which
distances the short distance behavior expected from the OPE sets in.
We also point out that a zero energy potential defined from the BS wave
function has a non-trivial form and may be a concept which may have a 
wider domain of applicability. 

We further examine how the wave function in the Ising model 
is built from the contributions of the intermediate particle states.
We find that intermediate states involving a relatively low number of 
particles give a good approximation down to quite short distances. 
This study was performed because in other integrable models the exact 
wave function is not (yet) known and the only analytical methods available 
are intermediate state approximations, the OPE (renormalized
perturbation theory at short distances) and $1/n$ expansions.
As an example in Sect.~3 we study the (asymptotically free) O(3) sigma 
model in two dimensions.

Various technical details are relegated to appendices and in 
Sect.~4 we make some concluding remarks.


\section{The two--dimensional Ising model in the scaling limit}

In this section we will discuss properties of BS wave functions
in the two--dimensional Ising field theory, but before introducing 
these we first briefly describe the theory 
and establish some notations and conventions.

The theoretical insight which is to be gained from the 2--d Ising 
model seems inexhaustible. In 1976 Wu, McCoy, Tracey and Barouch 
\cite{WMTB} showed that the model (at zero external field) has a 
continuum limit as one approaches the critical point. This  
relativistic quantum field theory, called the Ising field theory,
describes on--shell free particles of mass $M>0$. 
We denote the corresponding one--particle states with momentum 
$p=M(\cosh\theta,\sinh\theta)$ by $|\theta\rangle$, 
with state normalization
\be
\langle\theta'|\theta\rangle=4\pi\delta(\theta-\theta')\,.
\label{statenorm}
\end{equation}

The continuum limit of the spin field $\sigma(x)$ is an 
interpolating field for this particle; we chose the normalization
\be
\langle 0|\sigma(x)|\theta\rangle=\rme^{-ipx}\,,\,\,\,\,\,\,
p=M(\cosh\theta,\sinh\theta)\,.
\label{fieldnorm}
\end{equation}
Although the theory has an alternative representation in terms of free
(fermion) fields, the spin field is not a free field; nevertheless there
is a wealth of information on its correlation functions.
In ref.~\cite{WMTB} it was shown that the two--point function
of $\sigma(x)$ satisfies the Painlev\'{e} III equation. 
More explicitly defining the vacuum 2--point function
of $\sigma(x)$ and also that of the corresponding disorder variable $\mu(x)$ 
\footnote{which is local with respect to itself 
but non--local wrt $\sigma(x)$} at
equal times (which suffices for our considerations) by
\ba
G(r)&=&\langle 0|\sigma((0,x_1))\sigma(0)|0\rangle\,,
\\
\widetilde{G}(r)&=&\langle 0|\mu((0,x_1))\mu(0)|0\rangle\,,
\ea
where $r=Mx_1$. Then for the sum and difference
\be
G_\pm(r)=\widetilde{G}(r)\pm G(r)\,,
\label{Gplusminus}
\end{equation}
one has
\be
G_\pm(r)=\rme^{\chi(r)/2}\rme^{\pm\varphi(r)/2}\,,
\end{equation}
where the functions $\chi,\varphi$ obey the equations
\ba
\frac{1}{r}[r\varphi'(r)]'&=&\frac12\sh(2\varphi(r))\,,
\label{varphide}
\\
\frac{1}{r}[r\chi'(r)]'&=&\frac12[1-\ch(2\varphi(r))]\,.
\label{chide}\ea
For definiteness we are considering the theory obtained by taking the 
continuum limit from the symmetric phase, 
where $\mu(x)$ has a non--vanishing vacuum expectation value
\footnote{i.e. the fields $\sigma,\mu$ in Fonseca and 
Zamolodchikov \cite{FZ} are interchanged with respect to ours.
Also the field normalization differs.}. 

The short and long distance
behaviors of the functions $\varphi,\chi$ are summarized in Appendix~A, 
and from these it follows that for small $r>0$
(for the field normalization given in (\ref{fieldnorm})),
\be
G(r)\sim C_\chi r^{-1/4}+\rmO(r^{3/4}\ln r)\,,
\end{equation}
(where the constant $C_\chi$ is given in (\ref{Cchi}),) exhibiting the
well known anomalous dimension of $\sigma(x)$. For large 
distances $r>0$ the correlation function falls exponentially:
\be
G(r)\sim\frac{\rme^{-r}}{\sqrt{8\pi r}}\left[1+\rmO(r^{-1})\right]\,.
\end{equation}

The results on the 2--point function were subsequently derived 
in other ways (see e.g. \cite{SMJ} and \cite{BL}). 
A very elegant derivation recently presented by Fonseca
and Zamolodchikov \cite{FZ} is based on local conservation laws 
of the doubled Ising field theory. 

\subsection{Bethe--Salpeter wave functions}

In their remarkable paper Fonseca and Zamolodchikov \cite{FZ} showed that
their methods also lead to exact results for a larger class of 
correlation functions. In particular they obtained exact expressions 
for the BS wave functions for 2--particle in--states
\ba
\Psi(r,\theta)&=&
i\langle 0|\sigma((0,x_1))\sigma(0)|\theta,-\theta\rangle^{\rm in}\,,
\\
\widetilde{\Psi}(r,\theta)&=&
i\langle 0|\mu((0,x_1))\mu(0)|\theta,-\theta\rangle^{\rm in}\,.
\ea
In fact Fonseca and Zamolodchikov consider the wave functions 
for general space--time arguments of the fields but here we 
restrict ourselves to equal times. 
Without loss of generality we can consider the rapidity $\theta\ge0$ 
and in the following consider only $r>0$ since locality 
(and parity invariance) imply 
\be
\Psi(r,\theta)=\Psi(-r,\theta)\,.
\end{equation}

For the sum and difference
\be
\Psi_\pm=\widetilde{\Psi}\pm\Psi\,,
\end{equation}
Fonseca and Zamolodchikov \cite{FZ} obtain
\be
\Psi_\pm(r,\theta)=\frac{G_\pm(r)}{\ch\theta}\left[
\rme^{-\theta}\Phi_\pm(r,\theta)^2-\rme^{\theta}\Phi_\mp(r,\theta)^2\right]\,,
\end{equation}
where $G_\pm$ are defined in (\ref{Gplusminus}) and $\Phi_\pm$ 
\footnote{which are the functions 
$\Psi_\pm$ in the notation of ref.~\cite{FZ},} 
satisfy the coupled equations
\be
\Phi'_\pm(r,\theta)=\frac12\sh(\varphi(r)\pm\theta)\Phi_\mp(r,\theta)\,,
\label{Phipmde}
\end{equation}
and the boundary conditions for small $r$ are
\be
\rme^{\chi(r)/2}\Phi_\pm(r,\theta)\sim\sqrt{2\pi}C_\chi\rme^{\pm\theta/2}
r^{1/4}\left[1+\rmO(r^2\ln r)\right]\,.
\end{equation}

In terms of these functions the BS wave function $\Psi$ is given by
\be
\Psi(r,\theta)=\frac{\rme^{\chi(r)/2}}{\ch\theta}\left[
 \Phi_+(r,\theta)^2\cosh\left(\frac{\varphi(r)}{2}-\theta\right)
-\Phi_-(r,\theta)^2\cosh\left(\frac{\varphi(r)}{2}+\theta\right)\right]\,.
\end{equation}
For short distances $r$ it has the expansion
\ba
\Psi(r,\theta)&\sim&\Psi_{\rm as}(r,\theta)+\rmO(r^{7/4})\,,
\,\,\,\,\,
\\
\Psi_{\rm as}(r,\theta)&=&2\pi C_\chi r^{3/4}\sinh\theta\,,
\label{Psi_asympt}
\ea
which is as expected from the known operator expansion of 
the product of $\sigma$--fields
\be
\sigma((0,x_1))\sigma(0)\sim G(r)+cr^{3/4}{\cal E}(0)+....
\end{equation}
where ${\cal E}(x)$ is the mass operator of dimension 1.

\begin{figure}
\begin{center}
\psfig{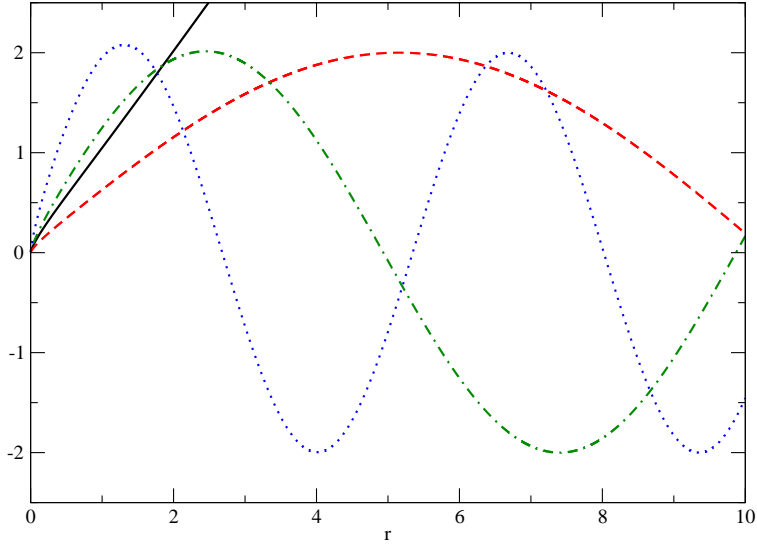}
\end{center}
\caption{\footnotesize The Ising BS wave function $\Psi(r,\theta)$ for 
$\theta=1.0$ (dotted), $\theta=0.6$ (dot-dashed), $\theta=0.3$ (dashed),
and the zero-energy wave function $\ell(r)$ defined in Appendix~B
(solid).}
\label{fig1a}
\end{figure}

\begin{figure}
\vspace{0.8cm}
\begin{center}
\psfig{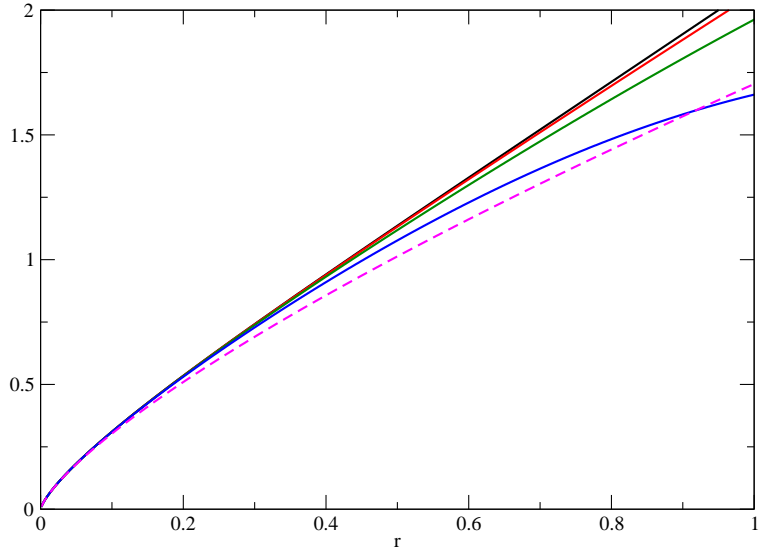}
\end{center}
\caption{\footnotesize A renormalized Ising BS wave function 
$\Psi(r,\theta)/\sinh(\theta)$ for 
for $\theta=0$ (top), $\theta=0.3,0.6,1.0$ (bottom solid curve).
The leading short distance OPE behavior $2\pi C_\chi r^{3/4}$ 
(see (\ref{Psi_asympt})) is given by the dashed curve.}
\label{fig1b}
\end{figure}

The coupled differential equations (\ref{varphide}), (\ref{chide}),
and (\ref{Phipmde}) for $\Phi_\pm, \varphi,\chi$ with their known 
boundary conditions at $r=0$ can be easily solved numerically. 
In Fig.~\ref{fig1a} we depict the wave function $\Psi(r,\theta)$ for 
various rapidities, illustrating the early set in of the 
long distance sinusoidal behavior.  
Fig.~\ref{fig1b} shows the wave function divided by $\sinh\theta$   
so that the leading short distance behaviors are the same,
(see (\ref{Psi_asympt})); once this renormalization is done there is
rather little remaining variation with the energy for $r<0.5$; 
moreover the leading OPE behavior dominates up to $r\sim0.2$.

\begin{figure}
\begin{center}
\psfig{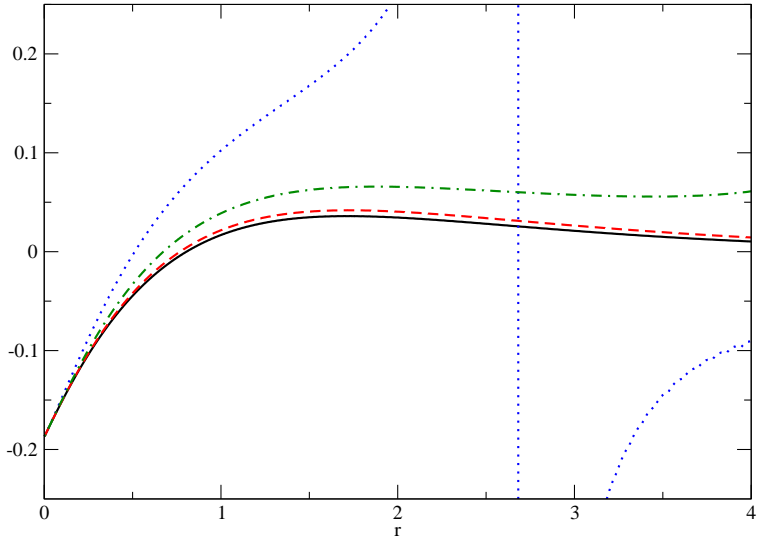}
\end{center}
\caption{\footnotesize The Ising BS potentials (multiplied by $r^2$) 
$r^2V_\theta(r)$ for $\theta=1.0$ (dotted), $\theta=0.6$ (dot-dashed), 
$\theta=0.3$ (dashed), and $\theta=0$ (solid).}
\label{fig2}
\end{figure}

\subsection{BS Potentials}

From the BS wave function one can define a rapidity--dependent potential by 
\be
V_\theta(r):=\frac{\Psi''(r,\theta)+\sinh^2\theta\Psi(r,\theta)}
{\Psi(r,\theta)}\,.
\label{BSpot}
\end{equation}
This definition is a direct analogy to that of energy dependent
NN potentials made in ref.~\cite{IAH1}. The hope is that for low energies
and for the distances relevant for phenomenology the potentials are only
mildly energy dependent
\footnote{Of course this excludes distances near and beyond the point 
where $\Psi(r,\theta)$ has its first zero 
and hence at which $V_\theta(r)$ is singular.}. 
It is such an ansatz which seems to qualitatively
apply in the NN case.  
We can investigate this question 
for the Ising field theory and find indeed only moderate variations 
in a reasonable range of parameters, as is illustrated in Fig.~\ref{fig2}.
The potential for $\theta=1.0$ becomes singular already at $r\sim2.681$
where the corresponding wave function has its first node.
Of course the physics in the Ising model is vastly different from the NN 
case, in particular in the Ising field theory there are no bound states.


The paper ref.~\cite{IAH2} describes some ideas to obtain a local energy
independent potential from the BS wave functions and this will hopefully 
be elucidated in our next paper \cite{ABHW}. 
Here we remark that a natural candidate 
for a potential of limited phenomenological relevance, 
as we will discuss below, is the zero--energy potential, 
obtained as the zero energy limit $V_0(r)$ of (\ref{BSpot}). 
For purposes of numerical evaluation in the Ising model this can 
be expressed in terms of $\varphi,\chi$ (see Appendix~A).
A plot of this potential is included in Fig.~\ref{fig2}. 
The asymptotic behaviors are analytically directly obtained from the
formulae in Appendix~A. From the large $r$ behavior of the zero energy 
wave function  
\begin{equation}
\Psi_0(r):=\lim_{\theta\to0}\left[\theta^{-1}\Psi(r,\theta)\right]\sim 
2r+\sqrt\frac{2}{\pi r}\rme^{-r}\left(1-\frac{17}{8r}+\dots\right)\,,
\label{zeroenwf}
\end{equation}
follows the leading large $r$ asymptotics of the zero energy potential:
\begin{equation}
V_0(r)\sim\frac{1}{\sqrt{2\pi}}\frac{1}{r^{3/2}}\,\rme^{-r}
\left(1-\frac{9}{8r}+\dots\right)\,,
\end{equation}
i.e. the potential falls exponentially to zero from above.
On the other hand for small $r$ using (\ref{Psi_asympt}) we have
\begin{equation}
V_0(r)\sim -\frac{3}{16}\frac{1}{r^2}\,.
\label{V0pot}
\end{equation}

Since this potential is (classically) strongly attractive close to the
origin the question of possible bound states naturally emerges. 
This would be fatal for the hope that the zero--energy potential is
at all relevant for the Ising field theory. In Appendix~A
we show that indeed there are no bound states because (\ref{V0pot})
is not attractive enough in the quantum theory.

In Appendix~B we consider the zero--energy potential in a slightly more 
general context. There we show that it reproduces the
correct scattering length, which parameterizes the leading low 
momentum behavior of the phase shift. However in general it does not
yield the exact next-to-leading behavior (although it may in some cases
be a good approximation to it).

\subsection{Intermediate particle state approximations to $\Psi(r,\theta)$}
 
In the Ising field theory we are, as discussed above, 
fortunate to have exact partial differential equations for the BS wave 
functions. However for most other integrable models 
this is not (yet) the case, and we have to resort to approximations
in order to obtain quantitative results. One approach is to compute 
contributions from intermediate states from knowledge of the form factors.
For the two--point function this approximation has been investigated  
in ref.~\cite{YZ}, and there the contributions of only a few states is 
found to approximate the exact result down to very small distances 
where the OPE can be applied.

In the Hilbert space of in--states defined by the spin field  
the S--matrix operator is given by
\be
{\bf S}=(-1)^{{\bf N}({\bf N}-1)/2}\,,
\end{equation}
where ${\bf N}$ is the particle number operator. An energy
independent phase is not observable in a scattering experiment; 
however the non-trivial S-matrix reflects the fact
that $\sigma(x)$ is not a free field.

Given knowledge of the S--matrix and assuming general properties 
such as analyticity and crossing symmetry (together with some additional
technical assumptions) it was argued in ref.~\cite{BKW} that generalized
form factors of the spin field are given by
\ba
&&\phantom{}^{\rm out}\langle\theta_1,\dots,\theta_t|\sigma(0)|
\theta_{t+1},\dots,\theta_n\rangle^{\rm in}
\nonumber\\
&&=(2i)^{(n-1)/2}\prod_{1\le i<j\le t}T(|\theta_i-\theta_j|)
\prod_{1\le r\le t<s\le n}\frac{\cal P}{T(\theta_r-\theta_s)}
\prod_{t<k<l\le n} T(|\theta_k-\theta_l|)\,,
\nonumber\\
&&
\ea
with $n$ an odd positive integer. ${\cal P}$ denotes the principle part 
and
\be
T(x)\equiv\tanh\frac{x}{2}\,.
\end{equation}

Sandwiching a complete set of states
\be
1=|0\rangle\langle 0|+\sum_{r=1}^\infty
\int_{-\infty}^\infty \frac{\rmd\theta_1}{4\pi}
\int_{-\infty}^{\theta_1}\frac{\rmd\theta_2}{4\pi}
\dots
\int_{-\infty}^{\theta_{r-1}}\frac{\rmd\theta_r}{4\pi}
|\theta_1,\dots,\theta_r\rangle^{\rm s}\phantom{}^{\rm s}
\langle \theta_1,\dots,\theta_r|
\end{equation}
(where s stands for in or out) between the fields, $\Psi$ can be expressed 
as a sum over $s$--particle contributions
\be
\Psi(r,\theta)=\sum_{s=1}^{\infty}\Psi_{2s-1}(r,\theta)\,.
\end{equation}

Starting with the 1--particle contribution we have
\be
\Psi_1(r,\theta)= -\frac{1}{2\pi}T(2\theta) p_1(r,\theta)\,.
\end{equation}
with
\be
p_1(r,\theta)=\int_{-\infty}^\infty\rmd\theta_1\,\rme^{ir\sh\theta_1}
\frac{{\cal P}}{T(\theta_1-\theta)T(\theta_1+\theta)}\,.
\end{equation}
Now use
\be
\frac{\cal P}{T(\phi)}=2\pi i\delta(\phi)+\frac{1}{T(\phi+i\epsilon)}\,,
\end{equation}
to obtain
\be
p_1=p_1^{(2)}+p_1^{(1)}+p_1^{(0)}\,,
\end{equation}
where the superscripts denote the number of delta functions. So
\ba
p_1^{(2)}(r,\theta)&=&(2\pi i)^2
\int_{-\infty}^{\infty}\rmd\theta_1\,\rme^{ir\sh\theta_1}
\delta(\theta_1-\theta)\delta(\theta_1+\theta)
\\
&=&-2\pi^2\delta(\theta)\,.
\ea
Next
\ba
p_1^{(1)}(r,\theta)&=&2\pi i
\int_{-\infty}^{\infty}\rmd\theta_1\,\rme^{ir\sh\theta_1}
\left[\frac{\delta(\theta_1-\theta)}{T(\theta_1+\theta+i\epsilon)}
     +\frac{\delta(\theta_1+\theta)}{T(\theta_1-\theta+i\epsilon)}\right]
\nonumber\\
&&
\\
&=&2\pi i\left[\frac{\rme^{ ir\sh\theta}}{T(2\theta+i\epsilon)}
              -\frac{\rme^{-ir\sh\theta}}{T(2\theta-i\epsilon)}
\right]
\\
&=&4\pi^2\delta(\theta)-4\pi\frac{\sin(r\sh\theta)}{T(2\theta)}\,.
\ea
Finally
\be
p_1^{(0)}(r,\theta)=\int_{-\infty}^\infty 
\rmd\theta_1\,\rme^{ir\sh\theta_1}
\frac{1}{T(\theta_1-\theta+i\epsilon)T(\theta_1+\theta+i\epsilon)}\,.
\end{equation}
Shifting the contour to the line parallel to the real axis with
imaginary part $i\pi/2$ (and observing that the contribution from the 
contours parallel to the imaginary axis at infinity is zero for 
$r\ne 0$) we get
\ba
p_1^{(0)}(r,\theta)&=&-\int_{-\infty}^\infty\rmd z\,
\rme^{-r\ch z}\left(\frac{\ch\theta+i\sh z}{\ch\theta-i\sh z}\right)
\\
&=&2K_0(r)-4\ch^2\theta\int_0^\infty\rmd z\,
\frac{\rme^{-r\ch z}}{\ch^2\theta+\sh^2 z}\,.
\ea
Summarizing we have
\be
\Psi_1(r,\theta)=
\frac{1}{2\pi}\left[4\pi\sin(r\sh\theta)
-T(2\theta)p_1^{(0)}(r,\theta)\right]\,.
\end{equation}
The plane wave part is as expected for a two particle S--matrix
equal to $-1$; and $p_1^{(0)}(r,\theta)$ decays exponentially
as $r\to\infty$. However $\Psi_1(r,\theta)$ diverges logarithmically  
as $r\to0$:
\be
\Psi_1(r,\theta)\sim \frac{1}{\pi}T(2\theta)
\left[\ln r+f(\theta)+\rmO(r)\right]\,,
\end{equation}
which is very different from the short distance behavior of the 
exact wave function given in (\ref{Psi_asympt}).

The contribution from the 3--particle states can be computed 
similarly. Here we just note that all contributions vanish
exponentially as $r\to\infty$ (some only as $\rme^{-r}$ 
due to disconnected contributions). We have numerically computed
these contributions for various rapidities and as a typical result 
we give the results for rapidity $\theta=0.3$ in Table~\ref{Isingwavefn03} 
where we compare the approximations to the exact wave function. 
We observe that whereas (for this rapidity) the 1--particle approximation 
fails quite badly at $r=0.1$, 
addition of the 3--particle contribution already
makes the agreement much better at this distance and already
here the asymptotic formula (\ref{Psi_asympt}) which can be derived 
from the OPE sets in. 
Addition of the 5--particle intermediate states would of course
improve the agreement to smaller distances as illustrated in
the O(3) $\sigma$--model in the next section.

\begin{table}[h]
\centering
\begin{tabular}[t]{l|l|l|l|l|l}
\hline
$r$ &
$\Psi_1(r,0.3)$\quad\,\,\,&
$\Psi_3(r,0.3)$\quad\,\,\,&  
$[\Psi_1+\Psi_3](r,0.3)$&  
$\Psi(r,0.3)$\quad\,\,\,&  
$\Psi_{\rm as}(r,0.3)$\\[1.0ex]
\hline \hline
$10.0$&$\phantom{-}1.92482\rme-1$&$1.39916\rme-6$&$\phantom{-}1.92484\rme-1$&$1.92484\rme-1$&\\[1.0ex]
$5.0$&$\phantom{-}1.99793$&$2.52957\rme-4$&$\phantom{-}1.99818$&$1.99818$&\\[1.0ex]
$4.0$&$\phantom{-}1.87759$&$7.18688\rme-4$&$\phantom{-}1.87831$&$1.87831$&\\[1.0ex]
$3.0$&$\phantom{-}1.58539$&$2.03727\rme-3$&$\phantom{-}1.58743$&$1.58743$&\\[1.0ex]
$2.0$&$\phantom{-}1.14970$&$5.73985\rme-3$&$\phantom{-}1.15544$&$1.15544$&\\[1.0ex]
$1.0$&$\phantom{-}6.12374\rme-1$&$1.63733\rme-2$&$\phantom{-}6.28747\rme-1$&$6.28748\rme-1$&\\[1.0ex]
$0.1$&$\phantom{-}2.60053\rme-4$&$9.40535\rme-2$&$\phantom{-}9.43136\rme-2$&$9.45618\rme-2$&$9.227\rme-2$\\[1.0ex]
$0.01$&$-2.43697\rme-1$&$2.50800\rme-1$&$\phantom{-}7.10370\rme-3$&$1.64495\rme-2$&$1.641\rme-2$\\[1.0ex]
$0.001$&$-4.59999\rme-1$&$3.85735\rme-1$&$-7.42642\rme-2$&$2.91863\rme-3$&$2.918\rme-3$\\[1.0ex]
$0.0001$&$ $&$ $&$ $&$5.18897\rme-4$&$5.189\rme-4$\\[1.0ex]
\hline
\end{tabular}
\caption{\footnotesize $\Psi(r,\theta)$ the exact wave function,
and $\Psi_1(r,\theta),\Psi_3(r,\theta)$ the 1-- and 3--particle
contributions, and the leading short distance behavior 
$\Psi_{\rm as}(r\theta)$ given in (\ref{Psi_asympt}), for $\theta=0.3$.}
\label{Isingwavefn03}
\end{table}

\vfill
\eject

\section{BS wave functions of the O(3) $\sigma$--model}

In this section we will give a quantitative discussion of 
BS wave functions and their associated potentials in the O(3)
non--linear sigma model in two dimensions. 
The O($n$) sigma model has long served as a favorite laboratory 
for testing ideas concerning asymptotically free theories 
\cite{Wolff90,HasMagNie,HasNie,LuWo,JanosMax}.
Unfortunately there is in this case
no exact expression known for correlation functions of any local operators. 
However for the case $n=3$ the 
multi--particle form factors (FF) (defined in Eq.~(\ref{sigFF})) 
can be obtained recursively (although they become extremely complicated
for higher particle states), and so one can often obtain excellent
approximations to correlation functions in a wide range of energies
by saturation with a low number of intermediate states. 

The spectrum is considered to contain an O($n$) vector multiplet of
particles of mass $M$ and to have no bound states.
The two-particle S--matrix established by Zamolodchikov and  
Zamolodchikov \cite{ZZ} is given by:
\be
S_{ab;cd}(\beta)=\sum_{I=0}^2 S_I(\beta)P_I(ab|cd)\,,
\label{Smatrix}
\end{equation}
where $\beta$ is the rapidity difference of the incoming particles
and $P_I$ are ``isospin" projectors given by
\footnote{Note
$\sum_{e,f}P_I(ab|ef)P_J(ef|cd)=\delta_{IJ}P_I(ab|cd)\,,$
and for $n=3$ one has $\sum_{a,b}P_I(ab|ab)=2I+1\,.$}
\begin{eqnarray}
P_0(ab\vert cd)&=&\frac{1}{n}\delta_{ab}\delta_{cd}\,,\\
P_1(ab\vert cd)&=&
\frac{1}{2}\delta_{ac}\delta_{bd}-\frac{1}{2}\delta_{ad}\delta_{bc}\,,\\
P_2(ab\vert cd)&=&
\frac{1}{2}\delta_{ac}\delta_{bd}+\frac{1}{2}\delta_{ad}\delta_{bc}
-\frac{1}{n}\delta_{ab}\delta_{cd}\,,\\
\nonumber
\end{eqnarray}
and
\begin{equation} 
S_I(\beta)=-(-1)^I\rme^{2i\delta_I(\beta)}\,.
\end{equation}
For the special case $n=3$ the phase shifts are simply given by
\ba
\delta_0(\beta)&=&-\arctan\left(\frac{\beta}{2\pi}\right)\,,
\\
\delta_2(\beta)&=&\arctan\left(\frac{\beta}{\pi}\right)\,,
\\
\delta_1(\beta)&=&\delta_0(\beta)+\delta_2(\beta)\,.
\ea

We define BS wave functions as in the last section by
\begin{equation} 
\Psi_{ab;cd}(x_1,\theta)=\langle0\vert\sigma^a(0,x_1)\sigma^b(0,0)\vert
c,\theta;d,-\theta\rangle^{{\rm in}},\qquad\qquad\theta>0\,.
\end{equation}
The spin field $\sigma^a(x)$ is an interpolating
field for the massive particle and we fix the normalization by
\be
\langle 0|\sigma^a(0)|b,\theta\rangle=\delta^{ab}\,.
\end{equation}
Translation invariance and locality implies
\begin{equation} 
\Psi_{ab;cd}(-x_1,\theta)=\Psi_{ba;dc}(x_1,\theta).
\label{loc}
\end{equation}

Introducing isospin components $T_I$ for all tensors $T_{ab;cd}\,$:
\begin{equation} 
T_I=\frac{1}{2I+1}P_I(ab\vert cd) T_{ab;cd}\,,
\end{equation}
Eq.~(\ref{loc}) now implies
\begin{equation} 
\Psi_I(-x_1,\theta)=(-1)^I\Psi_I(x_1,\theta)\,.
\label{loc2}
\end{equation}

As before the BS wave function can be expressed as an expansion over 
$s$--particle contributions:
\begin{equation} 
\Psi_{ab;cd}(x_1,\theta)=\sum_{s{\rm\,\,odd}}\Psi^{(s)}_{ab;cd}(x_1,\theta)\,,
\end{equation}
and these can further be organized in contributions having specific large
distance behavior. Since the computation is rather technical we relegate 
the details to Appendix~C and here just summarize the result.
Firstly it is convenient to introduce the modified wave function
\begin{equation} 
\widetilde\Psi_I(r,\theta)=\frac{i}{\tanh\theta}\rme^{-i\delta_I(2\theta)}
\Psi_I(r/M,\theta)\,,
\label{Psitilde}
\end{equation}
since as shown in Appendix~C $\widetilde\Psi$ becomes real (for real arguments).
This WF has a (large distance) expansion which is naturally
expressed in the form
\begin{equation}
\widetilde\Psi_I(r,\theta)=\sum_{m{\rm \ odd}} \left\{
A^{(m)}_I(r,\theta)+B^{(m)}_I(r,\theta)\right\}\,,
\label{exp1}
\end{equation}
where $A^{(m)}_I(r,\theta)\sim\rmO(\rme^{-mr})$
and $B^{(m)}_I(r,\theta)\sim\rmO(\rme^{-(m-1)r})$ for large $r$.
Their explicit expressions are given in Eqs.~(\ref{AIm}),(\ref{BIm})
and involve integrals over products of the generalized form factors
\be
\langle0|\sigma^a(0)|b_1,\beta_1;\dots;b_s,\beta_s\rangle^{\rm in}
={\cal F}^a_{b_1\dots b_s}(\beta_1,\dots,\beta_s)
\,,\,\,\beta_1>\beta_2>\dots\beta_s\,.
\label{sigFF}
\end{equation}
From the connectivity properties of the matrix elements built of
these form factors it follows 
that the $s$--particle contribution ($s\ge3$) contributes
not only to $A^{(s)}$ and $B^{(s)}$ but also to $A^{(s-2)}$. 
The leading term is
\begin{equation} 
B^{(1)}_I(r,\theta)=\frac{2}{\tanh\theta}(-1)^I\sin\left\{r\sinh\theta
+\delta_I(2\theta)\right\}\,.
\end{equation}

\begin{figure}
\begin{center}
\psfig{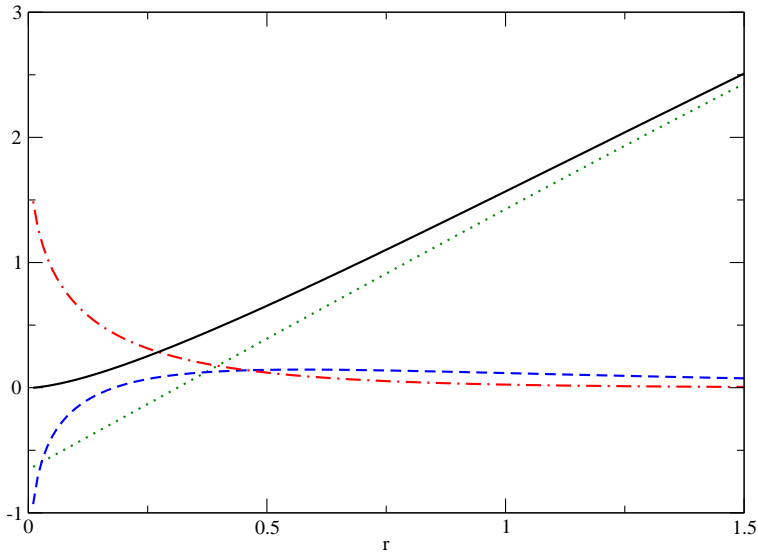}
\end{center}
\caption{\footnotesize Contributions to the O(3) wave function in the 
$I=0$ channel for $\theta=0.3$. 
The dotted curve is $B_0^{(1)}(r,0.3)$, the dashed curve is 
$A_0^{(1)}(r,0.3)$, the dot-dashed curve is $B_0^{(3)}(r,0.3)$,
and the solid curve is the sum of the first 5 contributions in the 
long distance expansion.}
\label{fig3}
\end{figure}

\begin{figure}
\begin{center}
\psfig{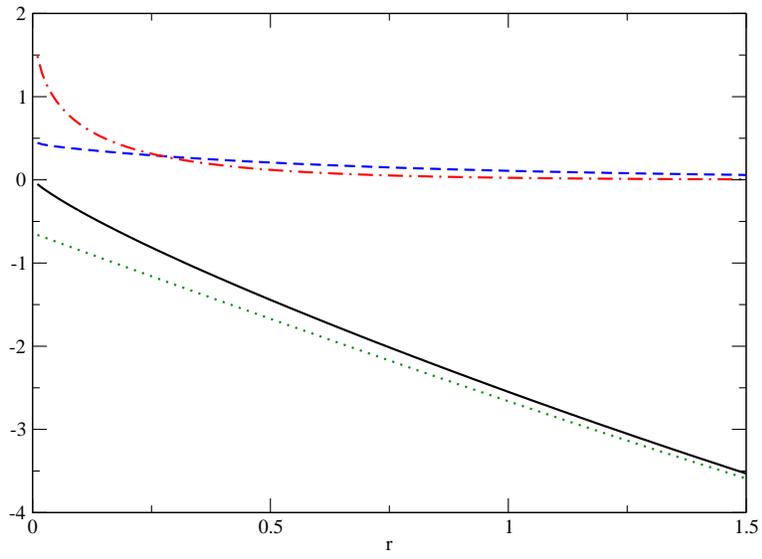}
\end{center}
\caption{\footnotesize As in Fig.~\ref{fig3} but for $I=1$.}
\label{fig4}
\end{figure}

\begin{figure}
\begin{center}
\psfig{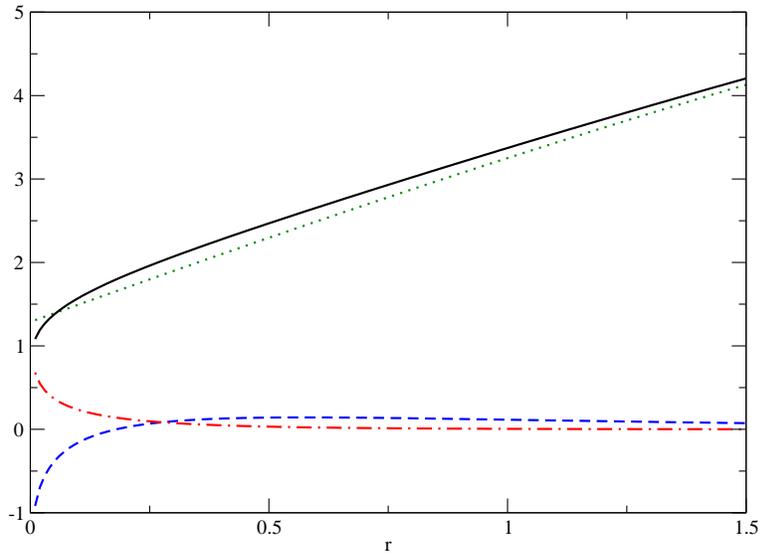}
\end{center}
\caption{\footnotesize As in Fig.~\ref{fig3} but for $I=2$.}
\label{fig5}
\end{figure}

We have numerically computed the first 5 terms in the expansion
for the three isospin values (in the case of O(3)) and a range of small 
($\le1.0$) rapidities.
As for the case of the Ising model, we find that inclusion of
sufficient number of terms in the long distance expansion gives
a good description of the full wave function down to quite small distances.
This is illustrated in Tables~\ref{O3I0wavefn0.3}--\ref{O3I2wavefn0.3}
in Appendix~D where we give results for $\theta=0.3$. 
Although the individual terms diverge at short distances it seems
that their sums are tending to zero in each channel see
Figs~\ref{fig3}--\ref{fig5}. Exactly how
the limit is reached can however not be read off from this approximation
and we require a detailed OPE analysis which we will present 
in the next subsection.

As in the case of the Ising models the BS wave functions and
their corresponding potentials are only
weakly varying with the rapidity (for moderate rapidities)
in the short to intermediate distance range.

Note for the zero energy WF we have
\begin{equation} 
B^{(1)}_I(r,0)=2(-1)^I\left\{r+a_I\right\}\,,
\end{equation}
where
\begin{equation} 
a_0=-\frac{1}{\pi},\qquad\quad
a_1=\frac{1}{\pi},\qquad\quad
a_2=\frac{2}{\pi}\,.
\end{equation}
For large $r$ the potentials fall exponentially to zero; in particular
for the zero energy potentials:
\ba
V_0(r)&=&\phantom{-}\frac{1}{18}\sqrt{\frac{2\pi^3}{r^3}}\left[1-0.615r^{-1}
+\rmO(r^{-2})\right]\rme^{-r}\,,
\\
V_1(r)&=&-\frac{1}{36}\sqrt{\frac{2\pi^3}{r^3}}\left[1-0.056 r^{-1}
+\rmO(r^{-2})\right]\rme^{-r}\,,
\\
V_2(r)&=&\phantom{-}\frac{1}{18}\sqrt{\frac{2\pi^3}{r^3}}\left[1-1.570r^{-1}
+\rmO(r^{-2})\right]\rme^{-r}\,.
\ea

\subsection{Operator product expansion}

Consider the operator product 
\begin{equation} 
D^{ab}_I(y)=P_I(ab\vert cd) \left(\sigma^c(y)\sigma^d(0)\right)\,,
\label{OPE}
\end{equation}
for general $n\ge3$. Using asymptotic freedom, we can show that its 
leading short distance expansion is of the form
\begin{equation} 
\begin{split}
D^{ab}_2(y)&\approx \alpha(\vert y\vert)\, t^{ab}(0)+\dots,\\
D^{ab}_1(y)&\approx \beta(\vert y\vert)\,y^\mu J^{ab}_\mu(0)+\dots,\\
D^{ab}_0(y)&\approx \frac{\delta^{ab}}{n}\Big\{\gamma_0(\vert y\vert)
+\vert y\vert^2\gamma_1(\vert y\vert)\Theta(0)
+y^\mu y^\nu\gamma_2(\vert y\vert)\widehat T_{\mu\nu}(0)\Big\}+\dots\\
\label{OPEofD}
\end{split}
\end{equation}
Here $t^{ab}$ is a traceless iso--tensor operator of dimension 0, 
$J^{ab}_\mu$ is the Noether current and $\Theta$ and $\widehat T_{\mu\nu}$ is 
the (Lorentz) trace and traceless part of the energy--momentum tensor 
$T_{\mu\nu}$:
\begin{equation} 
\Theta=T^\mu_{\phantom{\mu}\mu},\qquad\quad
T_{\mu\nu}=\widehat T_{\mu\nu}+\frac{1}{2}\eta_{\mu\nu}\Theta,
\end{equation}
$\eta_{00}=-\eta_{11}=1$. Finally the leading short distance behavior of 
the functions $\alpha(|y|),\beta(|y|),\gamma_j(|y|)$ appearing in
(\ref{OPEofD}) can be computed
in the framework of renormalized perturbation theory as discussed below.

Sandwiching the operator product between the vacuum and a two-particle
state we have
\begin{equation} 
\langle0\vert D^{ab}_I(y)\vert c,\theta;d,-\theta\rangle^{\rm in}=
P_I(ab\vert cd)\Psi_I(r,\theta)\,,
\end{equation}
where $r=M\vert y\vert$.

Let us recall \cite{CLUS} the two--particle form factors of the operators
occurring in the above short distance expansion:
\begin{equation} 
\langle0\vert t^{ab}(0)\vert c,\alpha;d,\beta\rangle^{\rm in}=
(\alpha-\beta-i\pi)\tanh\left(\frac{\alpha-\beta}{2}\right)P_2(ab\vert cd)\,,
\label{FF2}
\end{equation}
\begin{equation} 
\langle0\vert J^{ab}_\mu(0)\vert c,\alpha;d,\beta\rangle^{\rm in}=
-i\pi^2\epsilon_{\mu\nu}q^\nu\psi(\alpha-\beta)P_1(ab\vert cd)\,,
\end{equation}

\begin{equation} 
\langle0\vert T_{\mu\nu}(0)\vert a,\alpha;b,\beta\rangle^{\rm in}=
\frac{\pi^2}{2}\delta^{ab}(q^2\eta_{\mu\nu}-q_\mu q_\nu)
\frac{\psi(\alpha-\beta)}{\alpha-\beta-i\pi}\,.
\end{equation}
Here
\begin{equation} 
q^0=M(\cosh\alpha+\cosh\beta),\qquad\qquad
q^1=M(\sinh\alpha+\sinh\beta)\,,
\end{equation}
$\epsilon_{01}=-\epsilon_{10}=1$ and
\begin{equation} 
\psi(\theta)=\frac{\theta-i\pi}{\theta(2\pi i-\theta)}\tanh^2\left(
\frac{\theta}{2}\right)\,.
\label{psidef}
\end{equation}
Eq.~(\ref{FF2}) fixes the normalization of $t^{ab}$, which is otherwise
undetermined.

Using these form factors, the leading short distance expansion
of the isospin invariant wave functions are given as
\begin{equation} 
\begin{split}
\Psi_2(r,\theta)&\approx \alpha(r)(2\theta-i\pi)\tanh\theta\,,\\
\Psi_1(r,\theta)&\approx r\beta(r)\frac{i\pi^2\sinh\theta\tanh\theta}
{2\theta(i\pi-\theta)}(2\theta-i\pi)\,,\\
\Psi_0(r,\theta)&\approx (r\pi)^2[2\gamma_1(r)-\gamma_2(r)]
\frac{\sinh^2\theta}{4\theta(i\pi-\theta)}\,,\\
\end{split}
\end{equation}
which in terms of the redefined (real) field $\widetilde{\Psi}$ 
in (\ref{Psitilde}) read:
\begin{equation} 
\begin{split}
\widetilde\Psi_2(r,\theta)&\approx \alpha(r)\sqrt{\pi^2+4\theta^2}\,,\\
\widetilde\Psi_1(r,\theta)&\approx 
r\beta(r)\frac{\pi^2\sinh\theta}{2\theta}
\sqrt{\frac{\pi^2+4\theta^2}{\pi^2+\theta^2}}\,,\\
\widetilde\Psi_0(r,\theta)&\approx (r\pi)^2[2\gamma_1(r)-\gamma_2(r)]
\frac{\sinh2\theta}{8\theta}\frac{1}{\sqrt{\pi^2+\theta^2}}\,.\\
\end{split}
\end{equation}

We now outline the information which can be gained from 
perturbative field theory; for any undefined notation we refer the
reader to \cite{DIS}. We start with the short distance expansion
\begin{equation} 
\Delta^{ab}_I(y)=\frac{1}{g_0^2}P_I(ab\vert cd)S^c(y)S^d(0)\approx
\sum_\omega K^{(\omega)}_I(g_0,y){\cal B}^{(\omega)ab}_I(0)+\dots,
\label{OPEbare}
\end{equation}
where ${\cal B}^{(\omega)ab}_I$ are (bare) local operators and the
$K^{(\omega)}_I$ are coefficient functions (which can, in principle,
be calculated in perturbation theory). The operator product (\ref{OPE})
differs by a (non--perturbative) rescaling from the renormalized version
of (\ref{OPEbare}):
\begin{equation} 
\begin{split}
\Omega_n^{-2}D^{ab}_I(y)=\Delta^{ab}_{I(R)}(y)&=
P_I(ab\vert cd) S^c_{(R)}(y)S^d_{(R)}(0)\\
&\approx \sum_\omega k^{(\omega)}_I(g,\mu,y){\cal B}^{(\omega)ab}_{I(R)}(0)\,,\\
\end{split}
\end{equation}
written in terms of renormalized operators ${\cal B}^{(\omega)ab}_{I(R)}$
and finite coefficient functions $k^{(\omega)}_I$. The latter satisfies
the renormalization group (RG) equation
\begin{equation}
\left\{{\cal D}+\gamma(g)+\gamma^{(\omega)}_I(g)\right\}k^{(\omega)}_I=0\,, 
\label{RGeq}
\end{equation}
where the RG $\gamma$--function
\begin{equation} 
\gamma^{(\omega)}_I(g)=\gamma^{(\omega)}_{I0}g^2+\dots
\end{equation}
is related to the operator renormalization constant (in dimensional 
regularization) by
\begin{equation} 
Z^{(\omega)}_I=1-\frac{\gamma^{(\omega)}_{I0}g^2}{\varepsilon}+\dots,
\end{equation}
corresponding to the operator ${\cal B}^{(\omega)ab}_I\,$. 

In particular, for $I=2$ we have only one operator
\begin{equation} 
{\cal B}_2^{ab}=\tilde\tau^{ab}=\frac{1}{g_0^2}\left(S^aS^b-\frac{1}{n}
\delta^{ab}\right)\,,
\end{equation}
which has renormalization constant \cite{DIS}
\begin{equation} 
Z_2=Z_{\tilde\tau}=1+\frac{g^2}{\pi\varepsilon}+\dots
\end{equation}
and coefficient function 
\begin{equation} 
K_2(g_0,y)=1+\rmO(g_0^2)\,.
\end{equation}
Similarly, for $I=1$ we have
\begin{equation} 
B^{ab}_{1\mu}=J^{ab}_\mu\,,
\end{equation}
with $Z_1=1$ and
\begin{equation}
K^\mu_1(g_0,y)=-\frac{1}{2}y^\mu+\rmO(g_0^2)\,. 
\end{equation}
Finally for $I=0$ we have the two operators
\begin{equation}
{\cal B}_0^{(1)ab}=\frac{\delta^{ab}}{n}\Theta,\qquad\qquad
{\cal B}_{0\mu\nu}^{(2)ab}=\frac{\delta^{ab}}{n}\widehat T_{\mu\nu} 
\end{equation}
with $Z_0^{(\omega)}=1$ ($\omega=1,\,2$) and coefficient functions
\begin{equation} 
K_0^{(1)}=C_{10}\vert y\vert^2+\rmO(g_0^2),\qquad\qquad
K_0^{(2)\mu\nu}=-\frac{1}{2}y^\mu y^\nu+\rmO(g_0^2)\,.
\end{equation}
Since $\Theta$ vanishes at tree level, the corresponding numerical 
coefficient $C_{10}$ can only be determined by a one--loop calculation.

The equation (\ref{RGeq}) can be solved by standard RG methods. Introducing
the running coupling function $\lambda(r)$ as the solution of
\begin{equation} 
\frac{1}{\lambda(r)}+\chi\ln\lambda(r)=-\ln r\,, 
\label{lambda}
\end{equation}
with $\chi=1/(n-2)$ we find
\begin{equation} 
\begin{split}
\alpha(r)&\approx \alpha_0\,\lambda^\chi\left\{1+\rmO(\lambda)\right\}\,,\\
\beta(r)&\approx -\frac{1}{2}D_n\,(2\pi\chi\lambda)^{-\chi}
\left\{1+\rmO(\lambda)\right\}\,,\\
\gamma_1(r)&\approx C_{10}D_n\,(2\pi\chi\lambda)^{-\chi}
\left\{1+\rmO(\lambda)\right\}\,,\\
\gamma_2(r)&\approx -\frac{1}{2}D_n\,(2\pi\chi\lambda)^{-\chi}
\left\{1+\rmO(\lambda)\right\}\,.\\
\end{split}
\end{equation}
The constant occurring in the coefficient $\alpha$ cannot be calculated
since we do not know the relative normalization of the operator $\tilde
\tau^{ab}_{(R)}$ with respect to $t^{ab}$ whose normalization is fixed 
non-perturbatively by (\ref{FF2}). We also do not know the numerical value
of the non-perturbative constant $D_n$ for general $n$. However, we do know
\cite{DIS} $D_3=4/\pi$ and this enables us to write for $n=3$ (and, for 
simplicity, at zero energy):
\begin{equation} 
\begin{split}
\widetilde\Psi_0(r,0)&\approx 
\frac{r^2}{\pi\lambda}\left(C_{10}+\frac{1}{4}
\right)
\left\{1+c_0^{(1)}\lambda+\dots\right\}\,,\\
\widetilde\Psi_1(r,0)&\approx -\frac{r}{2\lambda}
\left\{1+c_1^{(1)}\lambda+\dots\right\}\,,\\
\widetilde\Psi_2(r,0)&\approx \alpha_0\pi\,\lambda 
\left\{1+c_2^{(1)}\lambda+\dots\right\}\,.\\
\end{split}
\label{3.46}
\end{equation}
Here the $\rmO(\lambda)$ (and higher) corrections can in principle be
calculated in higher orders of perturbation theory. The number
$C_{10}$ can also be obtained by a one--loop calculation. However, as
explained above, the coefficient $\alpha_0$ cannot be calculated by
presently available methods. 

Fortunately, the overall normalization cancels from the potential defined
by
\begin{equation} 
V_I(r)=\frac{\widetilde\Psi^{\prime\prime}_I(r,0)}{\widetilde\Psi_I(r,0)}\,,
\end{equation}
and we find
\begin{equation} 
\begin{split}
V_0(r)&\simeq \ \ \, 
\frac{2}{r^2}\left\{1-\frac32\lambda+\rmO(\lambda^2)\right\}\,,\\
V_1(r)&\simeq-\frac{\lambda}{r^2}
\left\{1+\left[1-c_1^{(1)}\right]\lambda+\rmO(\lambda^2)
\right\}\,,\\
V_2(r)&\simeq-\frac{\lambda}{r^2}
\left\{1-\left[1-c_2^{(1)}\right]\lambda+\rmO(\lambda^2)
\right\}\,.\\
\end{split}
\label{VOPE}
\end{equation}
Note that for $I=0$ we can calculate the first correction in
$\lambda(r)$ without knowledge of the $\rmO(\lambda)$ correction in 
(\ref{3.46}).

\begin{table}[h] 
\centering 
\begin{tabular}[t]{l|l|l|l|l} 
\hline 
$r$&$0.01$&$0.05$&$0.1$&$0.2$\\[1.0ex] 
\hline  
$\lambda$&$0.154$&$0.222$&$0.280$&$0.393$\\[1.0ex] 
\hline 
\end{tabular} 
\caption{\footnotesize Values of $\lambda(r)$.}  
\label{rtolambda} 
\end{table}

\begin{figure}
\begin{center}
\psfig{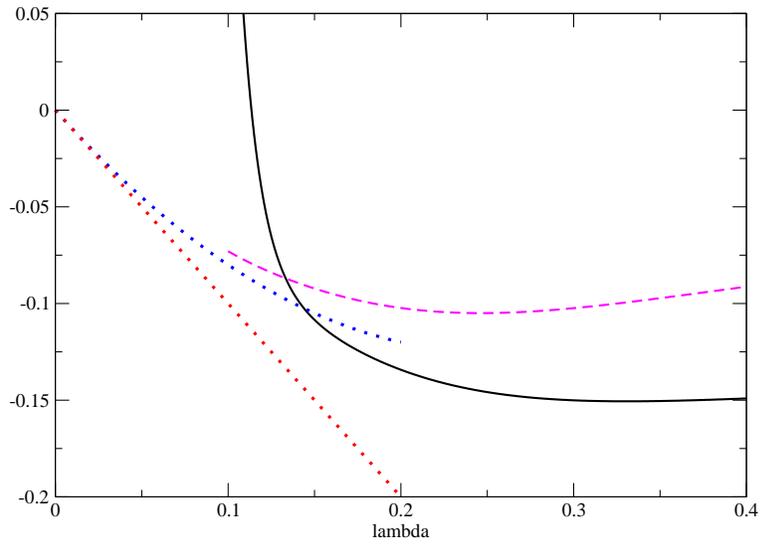}
\end{center}
\caption{\footnotesize ``Long distance approximation" to $r^2V_1(r)$ 
(solid) and $r^2V_2(r)$ (dashed). The lower dotted line is the
leading short distance behavior $-\lambda$ and the upper dotted line
is $-\lambda+2\lambda^2$.}
\label{fig6}
\end{figure}

In Fig.~\ref{fig6} we plot $r^2$ times the potentials in the
$I=1,2$ channels obtained from the sum of the first 5 leading 
terms in the long distance (LD) expansions, together with the leading 
behavior (\ref{VOPE}) obtained from the OPE. They are plotted 
with respect to the variable $\lambda$ defined in (\ref{lambda})
(with $\chi=1$); in Table~\ref{rtolambda} we give  
some pairs of values $(r,\lambda(r))$. We also plot the curve
$-\lambda+2\lambda^2$ to illustrate that 
(``quite reasonable") higher order expressions
in the OPE expansion could be found to make smooth meetings
with the LD approximations. We think that 
the 5--term LD approximation is accurate down to $\lambda\sim 0.2$ 
(which already corresponds to quite short distances) 
in the $I=1$ channel and even to smaller distances in the $I=2$ channel. 
This can be monitored by studying the stability of successive 
LD approximations, including only 3 terms, 4 terms and 5 terms 
respectively. This is illustrated in Fig.~\ref{fig6a} for $I=1$.
Alternativey one can appreciate the situation by inspecting
Table~\ref{O3zeropot} in Appendix~D where we give the double derivatives
of the separate contributions times $r^2$ in the various channels.

\begin{figure}
\begin{center}
\psfig{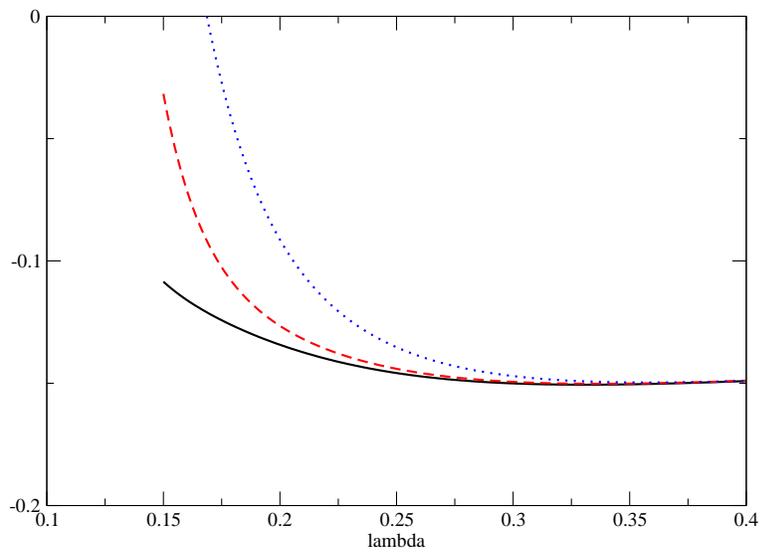}
\end{center}
\caption{\footnotesize Successive LD approximations to $r^2V_1(r)$; 
The dotted, dashed and solid lines correspond to approximations using
3,4,5 terms repectively.}
\label{fig6a}
\end{figure}

\begin{figure}
\begin{center}
\psfig{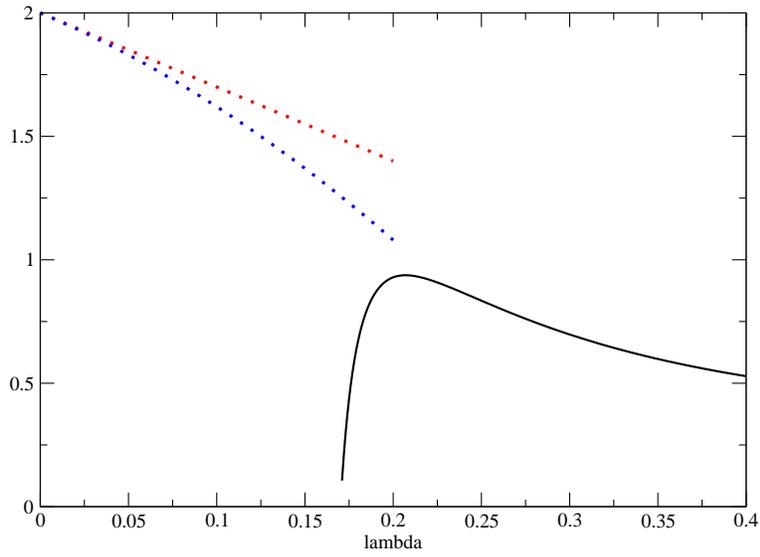}
\end{center}
\caption{\footnotesize LD approximation to $r^2V_0(r)$ (solid). 
The upper dotted line is the leading short distance behavior 
and the lower dotted line is $2-3\lambda-8\lambda^2$.}
\label{fig7}
\end{figure}

Fig.~\ref{fig7} shows the zero-energy potential in the $I=0$ channel.
Here the LD breaks down already larger distances, in fact 
there is no stability in the sense described above even at $r=0.1$.
The figure however suggests that our 5--term approximation may still be 
quite good there, but it would need computation of higher order terms 
to confirm this. Never the less it is plausible that the approximation
joins smoothly to the OPE expansion where again higher
order terms are required to improve the quantitative picture.

\vfill
\eject

\section{Conclusions}

In this paper we have investigated BS wave functions 
in integrable models; the Ising and O(3) $\sigma$--models. 
We have seen that potentials derived from them are rather slowly
varying with energy in the short and intermediate distance range
$M|x|<\sim1$. We have also discussed the relevance of the
zero-energy potential and its phenomenological limitations.
In these models we have found that a good approximation to the
wave functions can be obtained by combining a long distance expansion
(from contributions of intermediate states) and a short distance
expansion from the OPE. It would be instructive to study some
other examples in particular models with bound states.

Given a BS wave function constructed from a particular choice of 
local fields, it is clear that its short distance behavior 
can be obtained by an analysis of the OPE expansion.
Further it follows from naive dimensional analysis that the
BS potentials derived from (most) wave functions that vanish at 
the origin will behave as $|x|^{-2}$ (modified by
logarithms). The overall sign in 3--dimensions indicates
its attractive or repulsive nature. How this sign may depend on the 
particular interpolating field remains an important question.
  
In a sequel paper \cite{ABHW} we plan to include a more general discussion
on potentials obtained from BS wave functions. 
We will also present OPE predictions for the 
short distance behaviors of BS wave functions (and the resulting short 
distance behavior of the potentials) 
for the pion--pion and nucleon--nucleon cases in QCD, for some
choice of the interpolating fields..

\section*{Acknowledgments}

J.~B. and S.~A. are grateful to the Max--Planck--Institut f\"ur Physik for its
hospitality. This investigation was supported in part by the Hungarian
National Science Fund OTKA (under T049495) and the Grant-in-Aid of the
Japanese Ministry of Education (No. 20340047).

\vfill
\eject

\appendix
\renewcommand{\thesection}{Appendix~A}
\section{}
\renewcommand{\thesection}{A}

\subsection{Asymptotic behaviors of the functions $\varphi,\chi$}

Note here we always take $r>0$.
The short distance behaviors of $\varphi,\chi$ are given by \cite{FZ}:
\ba
\rme^{-\varphi(r)}&\sim&-\frac12 r\Omega(r)\left[1+\rmO(r^4)\right]\,, 
\\
\rme^{\chi(r)}&\sim&-2C_\chi^2\sqrt{r}\Omega(r)\left[1+\rmO(r^2)\right]\,, 
\ea
where
\be
\Omega(r)=\ln(kr)\,,\,\,\,\,\,\,\,k=\frac18\rme^\gamma\,,
\end{equation}
where $\gamma$ is the Euler--Mascheroni constant 	
($=0.57721566490\dots$), and
\be
C_\chi=2^{-7/6}A^{-3}\rme^{1/4}=0.27119012339\dots\,,
\label{Cchi}
\end{equation}
where $A$ is Glaisher's constant:
\be
A=\exp\left\{\frac{1}{12}-\zeta'(-1)\right\}=1.282427\dots
\end{equation}  

Next for long distances \cite{FZ}:
\ba
\varphi(r)&=&\frac{2}{\pi}K_0(r)+\rmO(\rme^{-3r})\,,
\\
\chi(r)&=&-2\ln2-\frac{2r}{\pi^2}\left[r\{K_0(r)^2-K_1(r)^2\}+
K_0(r)K_1(r)\right]+\rmO(\rme^{-4r})\,.
\ea
Recall the modified Bessel function 
\ba
K_0(r)&=&\int_0^\infty\rmd z\,\rme^{-r\ch z}\,,
\label{besselK0}
\ea
behaves for small $r$ as
\be
K_0(r)=-\ln\left(\frac{r}{2}\right)-\gamma+\rmO(r^2)\,,
\end{equation}
and for large $r$,
\be
K_0(r)=\sqrt{\frac{\pi}{2r}}\rme^{-r}
\left[1-\frac{1}{8r}+\dots\right]\,,
\end{equation}
and $K_1(r)=-K_0^\prime(r)\,$.

\subsection{Zero--energy potential}

Expanding $\Phi_\pm(r,\theta)$ for small $\theta$:
\be
\Phi_\pm(r,\theta)=\Phi_0(r)\pm\theta\Phi_1(r)+\dots
\end{equation}
and defining
\ba
\mu(r)&=&\Phi_0(r)^2\,,
\\
\nu(r)&=&2\Phi_0(r)\Phi_1(r)\,,
\ea
we have
\ba
\mu'(r)&=&\mu(r)\sinh\varphi(r)\,,
\\
\nu'(r)&=&\mu(r)\cosh\varphi(r)\,.
\ea
Then it follows for small $\theta$:
\be
\Psi(r,\theta)\sim 2\theta\rme^{\chi(r)/2}
\left[\nu(r)\cosh\frac{\varphi(r)}{2}
-\mu(r)\sinh\frac{\varphi(r)}{2}\right]+\dots
\end{equation}
and 
\ba
&&V_0(r):=\lim_{\theta\to0}\frac{\Psi''(r,\theta)}{\Psi(r,\theta)}
\nonumber\\
&&=\frac14\left[\chi'(r)\left(\frac{2}{r}+\chi'(r)\right)-\varphi'(r)^2\right]
+\frac{F(r)}
{\nu(r)\cosh\frac{\varphi(r)}{2}-\mu(r)\sinh\frac{\varphi(r)}{2}}\,.
\ea
with
\ba
&&F(r)=\frac12\left\{\chi'(r)\varphi'(r)+\varphi''(r)\right\}
\left\{\nu(r)\sinh\frac{\varphi(r)}{2}-\mu(r)\cosh\frac{\varphi(r)}{2}\right\}
\nonumber\\
&&+\mu(r)\left\{\chi'(r)+\sinh\varphi(r)\right\}\cosh\frac{\varphi(r)}{2}\,.
\ea

\subsection{Absence of bound states}

A bound state would have negative energy $-E$ and would be a solution
of the Schr\"odinger equation
\begin{equation}
-\Phi''+V_0\Phi=-E\Phi\,,
\end{equation}
with large $r$ asymptotics
\begin{equation}
\Phi(r)\sim\rme^{-\sqrt{E}r}\,, 
\end{equation}
i.e. would be normalizable.

It is important to impose the right boundary conditions at the origin.
It turns out \cite{LL,HN} that the correct boundary condition is
\begin{equation}
\Phi(r)\sim r^{3/4}\,.
\end{equation}
In other words, the other, more singular solution of the Schr\"odinger
equation, which would behave like
\begin{equation}
\Phi(r)\sim r^{1/4}
\end{equation}
is not allowed (the Hamilton operator would not be self-adjoint).

Using this information it is easy to derive the formula
\begin{equation}
\Psi'_0(r)\Phi(r)-\Psi_0(r)\Phi'(r)=-E\int_0^r\rmd x\,\Phi(x)\Psi_0(x)\,.
\end{equation}
The left hand side vanishes for large $r$ and we know that the zero
energy wave function $\Psi_0(r)$ (\ref{zeroenwf}) is positive everywhere. 
Therefore the bound state wave function has to change sign somewhere. 
We denote this point by $r_0$ and assume $\Phi(r)$ is positive between 
the origin and this point. Then we have
\begin{equation}
-\Psi_0(r_0)\Phi'(r_0)=-E\int_0^{r_0}\rmd x\,\Phi(x)\Psi_0(x)\,,
\end{equation}
implying that
\begin{equation}
\Phi'(r_0)>0\,,
\end{equation}
but this is obviously a contradiction, so $\Phi(r)$ cannot exist.
 
\vfill
\eject

\appendix
\renewcommand{\thesection}{Appendix~B. Scattering length and effective range}
\section{}
\renewcommand{\thesection}{B}


Let us consider the BS wave function
\begin{equation} 
\Phi(r,k)=\langle0\vert\sigma(0,r)\sigma(0,0)\vert2\rangle
\end{equation}
in a 2-dimensional model,
where $k$ is the wave number of the 2--particle state. Its large $r$ 
asymptotics
\begin{equation} 
\Phi(r,k)\sim \phi(k)\sin\left\{kr+\hat\delta(k)\right\}
\end{equation}
can be used \cite{BNNPSW} to read off the physical 
phase shift $\hat\delta(k)$. 
This has a low energy expansion of the form
\begin{equation} 
\hat\delta(k)=-\hat ak+\hat f k^3+\cdots
\end{equation}
It is convenient to introduce a new normalization for the wave function
so that for large $r$ we have
\begin{equation} 
\tilde\Phi(r,k)\sim\frac{\sin\left\{kr+\hat\delta(k)\right\}}{k}\,.
\end{equation}
The zero energy wave function
\begin{equation} 
\ell(r)=\tilde\Phi(r,0)\,,
\end{equation}
has long distance asymptotics
\begin{equation} 
\ell(r)\sim r-\hat a+\cdots
\end{equation}
It can be used to define the zero energy potential
\begin{equation} 
U(r)=\frac{\ell^{\prime\prime}(r)}{\ell(r)}\,.
\end{equation}

We can now study the solution of the Schr\"odinger equation with
this potential:
\begin{equation} 
-\psi^{\prime\prime}(r,k)+U(r)\psi(r,k)=k^2\psi(r,k)\,,
\label{Sch}
\end{equation}
where the solution is fixed by requiring the following asymptotic behavior:
\begin{equation} 
\psi(r,k)\sim \frac{\sin\left\{kr+\delta(k)\right\}}{k}\,,
\end{equation}
where for low energy
\begin{equation} 
\delta(k)=-ak+f k^3+\cdots
\end{equation}
Since the zero energy wave function is the same as before,
\begin{equation} 
\psi(r,0)=\ell(r)\,,
\end{equation}
we have
\begin{equation} 
a=\hat a\,,
\end{equation}
i.e. the scattering length corresponding to the zero energy potential
is equal to the physical one.

Using the Schr\"odinger equation (\ref{Sch}) we can write
\begin{equation} 
\frac{\rmd}{\rmd r}\left\{\psi^\prime(r,k_2)\psi(r,k_1)-
\psi^\prime(r,k_1)\psi(r,k_2)\right\}
=(k_1^2-k_2^2)\psi(r,k_1)\psi(r,k_2)\,,
\end{equation}
which can be integrated to give
\begin{equation} 
\psi^\prime(R,k_2)\psi(R,k_1)-
\psi^\prime(R,k_1)\psi(R,k_2)
=(k_1^2-k_2^2)\int_0^R\rmd r\,\psi(r,k_1)\psi(r,k_2)\,.
\label{psi}
\end{equation}

Now we define $S(r,k)$ by
\begin{equation} 
S(r,k):=\frac{\sin\left\{kr+\delta(k)\right\}}{k}\,,
\label{Skr}
\end{equation}
which satisfies
\begin{eqnarray} 
&&(k_1^2-k_2^2)\int_0^R\rmd r\,S(r,k_1) S(r,k_2)\nonumber\\
&=&S^\prime(R,k_2)S(R,k_1)-S^\prime(R,k_1)S(R,k_2)
+S^\prime(0,k_1)S(0,k_2)-S^\prime(0,k_2)S(0,k_1)\,.\nonumber\\
\label{S}
\end{eqnarray} 
Taking the difference between (\ref{S}) and (\ref{psi}) in the $R\to\infty$
limit gives
\begin{eqnarray} 
&&(k_1^2-k_2^2)\int_0^\infty\rmd r\,\left\{S(r,k_1) S(r,k_2)
-\psi(r,k_1)\psi(r,k_2)\right\}\nonumber\\
&=&\cos\delta(k_1)\frac{\sin\delta(k_2)}{k_2}
-\cos\delta(k_2)\frac{\sin\delta(k_1)}{k_1}\nonumber\\
\nonumber
\end{eqnarray} 
Now we take $k_2\to0$ and $k_1=k$ in the above formula and find
\begin{equation} 
k^2\int_0^\infty\rmd r\,\left\{(r-a) S(r,k)
-\psi(r,k)\ell(r)\right\}
=-a\cos\delta(k)-\frac{\sin\delta(k)}{k}\,.
\end{equation}
Expanding this exact formula we get
\begin{equation} 
f=\frac{a^3}{3}-B\,,
\end{equation}
where
\begin{equation} 
B=\int_0^\infty\rmd r\,\left\{(r-a)^2-\ell^2(r)\right\}\,.
\end{equation}

In scattering theory one usually introduces the effective range:
\begin{equation} 
\rho=\frac{B}{2a^2}\,,
\end{equation}
which, together with the scattering length $a$, gives a two--parameter
description of low energy scattering.

In the Ising model
\begin{equation} 
a=\hat a=0,\qquad\qquad \hat f=0\,,
\end{equation}
and $\rho$ cannot be defined. On the other hand
\footnote{The fact that $f>0$ follows directly from the property that $\ell(r)>r$ 
for all $r$.},
\begin{equation} 
f=-B=\int_0^\infty\rmd r\,\left\{\ell^2(r)-r^2\right\}\sim 0.263\,.
\end{equation}
We see that the zero energy potential does (in general) not reproduce the 
low energy expansion of the true phase shift beyond leading order.

\vfill
\eject

\appendix
\renewcommand{\thesection}{Appendix~C. $s$--particle 
contributions to the O(3) wave functions}
\section{}
\renewcommand{\thesection}{C}


The matrix elements built from the
form factors (\ref{sigFF})
have a particular connectivity structure (see e.g. ref.~\cite{Janos}),
from which one infers that 
the $s$--particle contribution can be written as a sum of three terms with
$D$ delta--functions involving the rapidity variables, $D=0,1,2$:
\begin{equation} 
\Psi^{(s)}_{ab;cd}(x_1,\theta)=\sum_{D=0}^2\Psi^{(s)(D)}_{ab;cd}(x_1,\theta)\,.
\end{equation}

We start with contributions without delta functions:
\begin{equation} 
\begin{split}
\Psi^{(s)(0)}_{ab;cd}(x_1,\theta)&=\int_{\beta_1>\cdots>\beta_s}
\frac{\rmd\beta_1\dots\rmd\beta_s}{(4\pi)^s}
\rme^{ix_1[\sinh\beta_1+\dots+\sinh\beta_s]}\\
&\times {\cal F}^a_{b_1\dots b_s}(\beta_1,\dots,\beta_s)
{\cal F}^b_{b_s\dots b_1cd}(\hat\beta_s,\dots,\hat\beta_1,\theta,-\theta)\,,\\
\end{split}
\end{equation}
where
\begin{equation} 
\hat\beta_i=\beta_i+i\pi-i\epsilon\,.
\end{equation}
We first note that here the integrand is a totally symmetric function
of the (real) integration variables $\beta_i$. Thus we can extend 
the integration from the original domain $\beta_1>\cdots>\beta_s$ to $\R^s$. 
After that we can shift the integration contour by defining
\begin{equation} 
\beta_i=\alpha_i-i\pi/2, \qquad\qquad \alpha_i{\rm \ real}\,.
\end{equation}
This latter step works for {\it negative} $x_1$ only, and for this 
reason from now on we take
\begin{equation}
Mx_1=-r,\qquad\qquad r>0\,, 
\end{equation}
and set $M=1$ in the following.
We will use (\ref{loc2}) later to get the wave function for positive $x_1$.
We get
\begin{equation} 
\Psi^{(s)(0)}_{ab;cd}(-r,\theta)=F^{(s)}_{ab;cd}(r,\theta),
\end{equation}
where
\begin{equation}
\begin{split}
F^{(s)}_{ab;cd}(r,\theta)&=\frac{1}{s!}\int_{-\infty}^\infty
\frac{\rmd \alpha_1\dots\rmd \alpha_s}{(4\pi)^s}
\rme^{-r[\cosh\alpha_1+\dots+\cosh \alpha_s]}\\
&\times {\cal F}^a_{b_1\dots b_s}(\alpha_1,\dots,\alpha_s)
{\cal F}^b_{b_s\dots b_1cd}(\alpha_s,\dots,\alpha_1,\theta-\frac{i\pi}{2},
-\theta-\frac{i\pi}{2})\,.\\
\end{split}
\label{Fn}
\end{equation}
This behaves as $\rmO(\rme^{-sr})$ for large $r$.

Next we discuss contributions with two delta-functions. Let us discuss first
the case $s=3$. Here we see using the FF axioms that the three terms have
the same analytic form, the only difference is that the range of the
integration variable $\beta$ is different, namely
\begin{equation} 
-\theta>\beta,\qquad\qquad\theta>\beta>-\theta,\qquad\qquad
\beta>\theta
\end{equation}
for the three terms. This means that the sum of the three contributions
can be simply written as the same integral with the integration
extending from $-\infty$ to $\infty$.

Similar considerations work for general $s$ and we get
\begin{equation} 
\begin{split}
\Psi^{(s)(2)}_{ab;cd}(x_1,\theta)&=\int_{\beta_3>\cdots>\beta_s}
\frac{\rmd\beta_3\dots\rmd\beta_s}{(4\pi)^{(s-2)}}
\rme^{ix_1[\sinh\beta_3+\dots+\sinh\beta_s]}\\
&\times {\cal F}^a_{cdb_3\dots b_s}(\theta,-\theta,\beta_3,\dots,\beta_s)
{\cal F}^b_{b_s\dots b_3}(\hat\beta_s,\dots,\hat\beta_3)\,.\\
\end{split}
\end{equation}
We can again extend the $\beta$ integrations to $\R^{(s-2)}$ and then
shift the integration contours:
\begin{equation} 
\begin{split}
\Psi^{(s)(2)}_{ab;cd}(-r,\theta)&=\frac{1}{(s-2)!}\int_{-\infty}^\infty
\frac{\rmd\alpha_3\dots\rmd\alpha_s}{(4\pi)^{(s-2)}}
\rme^{-r[\cosh\alpha_3+\dots+\cosh\alpha_s]}\\
&\times {\cal F}^a_{cdb_3\dots b_s}(\theta+\frac{i\pi}{2},
-\theta+\frac{i\pi}{2},\alpha_3,\dots,\alpha_s)
{\cal F}^b_{b_s\dots b_3}(\alpha_s,\dots,\alpha_3)\,.\\
\end{split}
\label{Psin2}
\end{equation}
Finally we use the relation (expressing parity invariance)
\begin{equation} 
{\cal F}^a_{b_1\dots b_s}(\theta_1,\dots,\theta_s)=
{\cal F}^a_{b_s\dots b_1}(-\theta_s,\dots,-\theta_1)\,,
\end{equation}
and see that (\ref{Psin2}) can be expressed with (\ref{Fn}):
\begin{equation} 
\Psi_{ab;cd}^{(s)(2)}(-r,\theta)=F^{(s-2)}_{ba;dc}(r,\theta)\,.
\end{equation}

The last group of integrals is with one delta-function. In the $s=3$ case
we can group the six contributions into two groups of three. The first
one contains integrals proportional to $\rme^{ix_1\sinh\theta}$, whereas
the integrals in the second one are proportional to 
$\rme^{-ix_1\sinh\theta}$ and also contain the factor 
$S_{cd;..}(2\theta)$.
Again, the three terms in the first group are of the same analytic form
and correspond to the domains
\begin{equation} 
\theta>\beta>\beta^\prime,\qquad\qquad\beta>\theta>\beta^\prime,\qquad\qquad
\beta>\beta^\prime>\theta\,,
\end{equation}
for the two integration variables $\beta,\beta^\prime$. The sum of the
three integrals is simply an integral over $\beta>\beta^\prime$.
Generalizing to arbitrary $s$ the first group gives
\begin{equation} 
\begin{split}
\Psi^{(s)(1){\rm first}}_{ab;cd}(x,\theta)&=\rme^{ix_1\sinh\theta}
\int_{\beta_2>\cdots>\beta_s}
\frac{\rmd\beta_2\dots\rmd\beta_s}{(4\pi)^{(s-1)}}
\rme^{ix_1[\sinh\beta_2+\dots+\sinh\beta_s]}\\
&\times {\cal F}^a_{cb_2\dots b_s}(\theta,\beta_2,\dots,\beta_s)
{\cal F}^b_{b_s\dots b_2d}(\hat\beta_s,\dots,\hat\beta_2,-\theta)\,.\\
\end{split}
\end{equation}
Performing the usual operations we get
\begin{equation} 
\Psi_{ab;cd}^{(s)(1)}(-r,\theta)=
\rme^{-ir\sinh\theta}g^{(s)}_{ab;cd}(r,\theta)+
\rme^{ir\sinh\theta}S_{cd;d^\prime c^\prime}(2\theta)
g^{(s)}_{ab;c^\prime d^\prime}(r,-\theta)\,,
\end{equation}
where
\begin{equation} 
\begin{split}
g^{(s)}_{ab;cd}(r,\theta)&=\frac{1}{(s-1)!}\int_{-\infty}^\infty
\frac{\rmd\alpha_2\dots\rmd\alpha_s}{(4\pi)^{(s-1)}}
\rme^{-r[\cosh\alpha_2+\dots+\cosh\alpha_s]}\\
&\times {\cal F}^a_{cb_2\dots b_s}(\theta+\frac{i\pi}{2},\alpha_2,\dots,\alpha_s)
{\cal F}^b_{b_s\dots b_2d}(\alpha_s,\dots,\alpha_2,-\theta-\frac{i\pi}{2})\,.\\
\end{split}
\label{gn}
\end{equation}
It is easy to see that this behaves as $\rmO(\rme^{-(s-1)r})$ for large 
$r$.

Adding up all the contributions and using (\ref{loc2}) we can write for
the isospin components:
\begin{equation}
\begin{split} 
\Psi_I(r,\theta)&=2(-1)^I\sum_{s{\rm \ odd}} F^{(s)}_I(r,\theta)\\
&+\sum_{s{\rm \ odd}}\left\{(-1)^I\rme^{-ir\sinh\theta}
g^{(s)}_I(r,\theta)+\rme^{ir\sinh\theta}
S_I(2\theta)g^{(s)}_I(r,-\theta)\right\}\,,\\
\end{split}
\label{exp}
\end{equation}
where $S_I$ are the isospin invariant S--matrix amplitudes in 
(\ref{Smatrix}). Equation (\ref{exp}) is a large distance expansion 
and we now study the first few 
terms (up to the ones behaving $\rmO(\rme^{-3r})$ for large $r$).

For $\rmO(1)$ we get
\begin{equation} 
g^{(1)}_I(r,\theta)=1\,.
\end{equation}
For $\rmO(\rme^{-r})$ we have
\begin{equation} 
F^{(1)}_I(r,\theta)=\frac{\pi^2}{4}\psi(2\theta)\int_{-\infty}^\infty
\rmd\alpha\rme^{-r\cosh\alpha}\psi(\alpha+\frac{i\pi}{2}-\theta)
\psi(\alpha+\frac{i\pi}{2}+\theta)\rho_I(\alpha,\theta),
\end{equation}
where
\begin{equation} 
\rho_0(\alpha,\theta)=-4\theta-2\pi i\,,\qquad
\rho_1(\alpha,\theta)=i\pi-2\alpha\,,\qquad
\rho_2(\alpha,\theta)=2\theta-2\pi i\,.
\end{equation}
Here we have used the representation (for $s\ge2$):
\be
{\cal F}^a_{b_1\dots b_s}(\theta_1,\dots,\theta_s)
=\pi^{3(s-1)/2}g^a_{b_1\dots b_s}(\theta_1,\dots,\theta_s)
\prod_{1\le i<j\le s}\psi(\theta_i-\theta_j)\,,
\end{equation}
with $\psi(\theta)$ defined in (\ref{psidef})
and where $g^a_{b_1\dots b_s}(\theta_1,\dots,\theta_s)$ is a polynomial
function of the rapidities. Some explicit expressions of the latter 
are given in \cite{JanosMax}.

Next we have
\begin{equation} 
\begin{split}
g^{(3)}_I(r,\theta)&=\frac{\pi^4}{32}\int_{-\infty}^\infty
\rmd\alpha_2\rmd\alpha_3\rme^{-r(\cosh\alpha_2+\cosh\alpha_3)}
h_I(\alpha_2,\alpha_3,\theta)\\
&\times \psi(\alpha_2-\alpha_3)\psi(\alpha_3-\alpha_2)
\psi(\theta+\frac{i\pi}{2}-\alpha_2)
\psi(\theta+\frac{i\pi}{2}-\alpha_3)\\
&\times \qquad\psi(\theta+\frac{i\pi}{2}+\alpha_2)
\psi(\theta+\frac{i\pi}{2}+\alpha_3)\,,\\
\end{split}
\end{equation}
where $h_I$ is the quadratic polynomial
\begin{equation} 
h_I(\alpha_2,\alpha_3,\theta)=\frac{1}{2I+1}P_I(ab\vert cd)
g^a_{cb_2b_3}(\theta+\frac{i\pi}{2},\alpha_2,\alpha_3)
g^b_{b_3b_2d}(\alpha_3,\alpha_2,-\theta-\frac{i\pi}{2})\,.
\end{equation}
Finally
\begin{equation} 
\begin{split}
F^{(3)}_I(r,\theta)&=\frac{\pi^6}{384}\psi(2\theta)\int_{-\infty}^\infty
\rmd\alpha_1\rmd\alpha_2\rmd\alpha_3
\rme^{-r(\cosh\alpha_1+\cosh\alpha_2+\cosh\alpha_3)}
\omega_I(\alpha_1,\alpha_2,\alpha_3,\theta)\\
&\times\prod_{k=1}^3 \psi(\alpha_k+\frac{i\pi}{2}+\theta)
\psi(\alpha_k+\frac{i\pi}{2}-\theta)
\prod_{k<l}\psi(\alpha_k-\alpha_l)\psi(\alpha_l-\alpha_k)
\end{split}
\end{equation}
with
\begin{equation} 
\begin{split}
\omega_I(\alpha_1,\alpha_2,\alpha_3,\theta)&=\frac{1}{2I+1}P_I(ab\vert cd)
g^a_{b_1b_2b_3}(\alpha_1,\alpha_2,\alpha_3)\\
&\times g^b_{b_3b_2b_1cd}(\alpha_3,\alpha_2,\alpha_1,\theta-\frac{i\pi}{2},-\theta-\frac{i\pi}{2}).
\end{split}
\end{equation}

Let us now study the phase of the WF. We start from the relation
\begin{equation} 
\left\{{\cal F}^a_{b_1\dots b_s}(\theta_1,\dots,\theta_s)\right\}^*=
{\cal F}^a_{b_1\dots b_s}(-\theta_1^*,\dots,-\theta_s^*),
\label{real}
\end{equation}
expressing the fact that the FF is a real analytic function (which is
a consequence of CPT symmetry, but can also be proven directly). 
Using this in (\ref{Fn}) we find the relation
\begin{equation} 
\left\{F^{(s)}_{ab;cd}(r,\theta)\right\}^*=
S_{cd;yx}(-2\theta)F^{(s)}_{ab;xy}(r,\theta)\,,
\end{equation}
which gives
\begin{equation} 
\left\{F^{(s)}_I(r,\theta)\right\}^*=(-1)^I
S_I(-2\theta)F^{(s)}_I(r,\theta)=
-\rme^{-2i\delta_I(2\theta)}F^{(s)}_I(r,\theta)\,.
\label{cc1}
\end{equation}
For (\ref{gn}) we simply get
\begin{equation} 
\left\{g^{(s)}_{ab;cd}(r,\theta)\right\}^*=
g^{(s)}_{ab;cd}(r,-\theta)\,,
\end{equation}
and
\begin{equation} 
\left\{g^{(s)}_I(r,\theta)\right\}^*=g^{(s)}_I(r,-\theta).
\label{cc2}
\end{equation}
Using (\ref{exp}) and the relations (\ref{cc1}), (\ref{cc2}) we see
that $\widetilde{\Psi}$ defined in (\ref{Psitilde}) is real.
Further the functions occurring in its long distance expansion
(\ref{exp1}) are given by
\begin{equation} 
A^{(m)}_I(r,\theta)=\frac{2i}{\tanh\theta}(-1)^I\rme^{-i\delta_I(2\theta)}
F^{(m)}_I(r,\theta)\,,
\label{AIm}
\end{equation}
and
\begin{equation} 
B^{(m)}_I(r,\theta)=\frac{-2}{\tanh\theta}(-1)^I{\rm Im}
\left\{
\rme^{-ir\sinh\theta}\rme^{-i\delta_I(2\theta)}
g^{(m)}_I(r,\theta)\right\}\,,
\label{BIm}
\end{equation}
with $F^{(m)},g^{(m)}$ given in Eqs.~(\ref{Fn}), and (\ref{gn})
respectively.

\vfill
\eject

\appendix
\renewcommand{\thesection}{Appendix~D. O(3) $\sigma$ model tables}
\section{}
\renewcommand{\thesection}{D}

\newcommand{\phm}{\phantom{-}}
\begin{table}[h]
\centering 
\begin{tabular}[t]{l|l|l|l|l|l|l} 
\hline 
$r$ 
&$B_0^{(1)}(r,0.3)$
&$A_0^{(1)}(r,0.3)$
&$B_0^{(3)}(r,0.3)$
&$A_0^{(3)}(r,0.3)$
&$B_0^{(5)}(r,0.3)$
&sum\\[1.0ex] 
\hline \hline 
$10.0$&$1.30735$&$1.06995\rme-5$&$-8.766\rme-12$&$8.0801\rme-21$
&$-3.1\rme-28$&$1.30736$\\[1.0ex]
$ 5.0$&$6.79501$&$1.96768\rme-3$&$6.5182\rme-7$&$2.6482\rme-13$
&$2.98\rme-18$&$6.79698$\\[1.0ex]
$ 4.0$&$6.18820$&$5.6340\rme-3$&$8.8594\rme-6$&$1.0298\rme-11$
&$1.40\rme-15$&$6.19384$\\[1.0ex]
$ 3.0$&$5.01196$&$1.61265\rme-2$&$1.1513\rme-4$&$4.5395\rme-10$
&$5.51\rme-13$&$5.02820$\\[1.0ex]
$ 2.0$&$3.37453$&$4.55362\rme-2$&$1.5472\rme-3$&$2.4111\rme-8$
&$2.86\rme-10$&$3.42161$\\[1.0ex]
$ 1.0$&$1.42658$&$0.116843$&$2.4630\rme-2$&$1.6800\rme-6$
&$3.34\rme-7$&$1.56806$\\[1.0ex]
$ 0.1$&$-0.444244$&$-0.162087$&$0.666238$&$-1.3290\rme-3$
&$5.355\rme-3$&$0.063932$\\[1.0ex]
$ 0.01$&$-0.631821$&$-0.930030$&$1.48854$&$-4.1388\rme-2$
&$0.11430$&$-0.000397$\\[1.0ex]
$ 0.001$&$-0.650555$&$-1.53211$&$1.93324$&$-0.21387$
&$0.44076$&$-0.02252$\\[1.0ex]
$ 0.0001$&$-0.652428$&$-1.93408$&$2.12862$&$-0.54207$
&$0.9129$&$-0.0870$\\[1.0ex]
\hline 
\end{tabular} 
\caption{\footnotesize $s=1,3$ and $B^{(5)}$ contributions
to O(3) isospin 0 wave functions for $\theta=0.3$.}  
\label{O3I0wavefn0.3} 
\end{table}

\begin{table} 
\centering 
\begin{tabular}[t]{l|l|l|l|l|l|l} 
\hline 
$r$
&$B_1^{(1)}(r,0.3)$
&$A_1^{(1)}(r,0.3)$
&$B_1^{(3)}(r,0.3)$
&$A_1^{(3)}(r,0.3)$
&$B_1^{(5)}(r,0.3)$
&sum\\[1.0ex] 
\hline \hline 
$10.0$&$-0.00197770$&$5.8085\rme-6$&$-1.84\rme-12$&$-7.4550\rme-21$
&$-1.0\rme-29$&$-0.00197711$\\[1.0ex]
$ 5.0$&$-6.85843$&$1.1598\rme-3$&$5.9610\rme-8$&$-2.7141\rme-13$
&$1.41\rme-19$&$-6.85727$\\[1.0ex]
$ 4.0$&$-6.63613$&$3.4459\rme-3$&$1.0439\rme-6$&$-1.1041\rme-11$
&$5.16\rme-17$&$-6.63268$\\[1.0ex]
$ 3.0$&$-5.80318$&$1.0444\rme-2$&$1.5554\rme-5$&$-5.209\rme-10$
&$1.82\rme-14$&$-5.79272$\\[1.0ex]
$ 2.0$&$-4.43624$&$3.2680\rme-2$&$2.2998\rme-4$&$-3.1147\rme-8$
&$8.44\rme-12$&$-4.40333$\\[1.0ex]
$ 1.0$&$-2.66108$&$0.108668$&$3.9490\rme-3$&$-2.9371\rme-6$
&$8.31\rme-9$&$-2.54846$\\[1.0ex]
$ 0.1$&$-0.848874$&$0.368994$&$0.102435$&$-2.7406\rme-4$
&$7.407\rme-5$&$-0.37764$\\[1.0ex]
$ 0.01$&$-0.661862$&$0.445499$&$0.157496$&$6.0312\rme-3$
&$1.528\rme-3$&$-0.05131$\\[1.0ex]
$ 0.001$&$-0.643131$&$0.524720$&$7.8661\rme-2$&$2.3673\rme-2$
&$9.049\rme-3$&$-0.00703$\\[1.0ex]
$ 0.0001$&$-0.641257$&$0.621496$&$-0.054472$&$0.046377$
&$0.024426$&$-0.00343$\\[1.0ex]
\hline 
\end{tabular} 
\caption{\footnotesize As in Table~\ref{O3I0wavefn0.3}
but for isospin 1.}  
\label{O3I1wavefn0.3} 
\end{table}

\begin{table} 
\centering 
\begin{tabular}[t]{l|l|l|l|l|l|l} 
\hline 
$r$ 
&$B_2^{(1)}(r,0.3)$
&$A_2^{(1)}(r,0.3)$
&$B_2^{(3)}(r,0.3)$
&$A_2^{(3)}(r,0.3)$
&$B_2^{(5)}(r,0.3)$
&sum\\[1.0ex] 
\hline \hline 
$10.0$&$-0.632946$&$1.0557\rme-5$&$-2.03\rme-12$&$1.211\rme-20$
&$-1.9\rme-30$&$-0.632935$\\[1.0ex]
$ 5.0$&$6.79781$&$1.9415\rme-3$&$5.5835\rme-8$&$4.6314\rme-13$
&$-1.06\rme-19$&$6.79975$\\[1.0ex]
$ 4.0$&$6.77335$&$5.5592\rme-3$&$1.1078\rme-6$&$1.9250\rme-11$
&$-1.35\rme-17$&$6.77891$\\[1.0ex]
$ 3.0$&$6.12563$&$1.5912\rme-2$&$1.7834\rme-5$&$9.3827\rme-10$
&$-7.34\rme-16$&$6.14156$\\[1.0ex]
$ 2.0$&$4.91424$&$4.4931\rme-2$&$2.8628\rme-4$&$5.9387\rme-8$
&$1.19\rme-12$&$4.95945$\\[1.0ex]
$ 1.0$&$3.25064$&$0.115291$&$5.6063\rme-3$&$6.4031\rme-6$
&$3.37\rme-9$&$3.37155$\\[1.0ex]
$ 0.1$&$1.49266$&$-0.159933$&$0.235546$&$2.5843\rme-3$
&$9.673\rme-5$&$1.57095$\\[1.0ex]
$ 0.01$&$1.30846$&$-0.917675$&$0.680703$&$1.10920\rme-2$
&$2.688\rme-4$&$1.08285$\\[1.0ex]
$ 0.001$&$1.28998$&$-1.51175$&$1.03886$&$3.0193\rme-2$
&$-0.010478$&$0.83681$\\[1.0ex]
$ 0.0001$&$1.28814$&$-1.90839$&$1.27147$&$7.1185\rme-2$
&$-0.037702$&$0.68470$\\[1.0ex]
\hline 
\end{tabular} 
\caption{\footnotesize As in Table~\ref{O3I0wavefn0.3}
but for isospin 2.}  
\label{O3I2wavefn0.3} 
\end{table}

\begin{table}
\centering 
\begin{tabular}[t]{l|l|l|l|l|l} 
\hline 
$r$ 
&$r^2A_0^{(1)''}(r,0)$
&$r^2B_0^{(3)''}(r,0)$
&$r^2A_0^{(3)''}(r,0)$
&$r^2B_0^{(5)''}(r,0)$
&$r^2V_0(r)$\\[1.0ex] 
\hline \hline 
$5.0$&$5.312\rme-2$&$2.60\rme-4$&$8.86\rme-11$&$2.26\rme-14$&$5.70\rme-3$\\[1.0ex]
$2.0$&$0.16415$&$5.273\rme-2$&$1.66\rme-6$&$7.99\rme-8$&$6.36\rme-2$\\[1.0ex]
$1.0$&$-2.875\rme-2$&$0.24430$&$3.10\rme-5$&$2.71\rme-5$&$0.1436$\\[1.0ex]
$0.2$&$-0.39396$&$0.47923$&$-2.88\rme-3$&$1.035\rme-2$&$0.54*$\\[1.0ex]
$0.1$&$-0.39508$&$0.41926$&$-9.72\rme-3$&$3.086\rme-2$&$**$\\[1.0ex]
\hline\hline 
$r$ 
&$r^2A_1^{(1)''}(r,0)$
&$r^2B_1^{(3)''}(r,0)$
&$r^2A_1^{(3)''}(r,0)$
&$r^2B_1^{(5)''}(r,0)$
&$r^2V_1(r)$\\[1.0ex] 
\hline \hline 
$5.0$&$3.439\rme-2$&$4.66\rme-5$&$-8.98\rme-11$&$5.94\rme-16$&$-3.24\rme-3$\\[1.0ex]
$2.0$&$0.18153$&$9.74\rme-3$&$-2.27\rme-6$&$1.77\rme-9$&$-4.15\rme-2$\\[1.0ex]
$1.0$&$0.17885$&$4.503\rme-2$&$-7.28\rme-5$&$5.03\rme-7$&$-8.87\rme-2$\\[1.0ex]
$0.1$&$1.184\rme-2$&$4.100\rme-2$&$1.54\rme-3$&$3.26\rme-4$&$-0.1490$\\[1.0ex]
$0.05$&$1.275\rme-2$&$1.135\rme-2$&$3.761\rme-3$&$7.17\rme-4$&$-0.141*$\\[1.0ex]
$0.01$&$3.256\rme-2$&$-3.763\rme-2$&$7.74\rme-3$&$2.92\rme-3$&$**$\\[1.0ex]
\hline\hline 
$r$ 
&$r^2A_2^{(1)''}(r,0)$
&$r^2B_2^{(3)''}(r,0)$
&$r^2A_2^{(3)''}(r,0)$
&$r^2B_2^{(5)''}(r,0)$
&$r^2V_2(r)$\\[1.0ex] 
\hline \hline 
$5.0$&$5.312\rme-2$&$5.52\rme-5$&$1.55\rme-10$&$3.16\rme-16$&$4.72\rme-3$\\[1.0ex]
$2.0$&$0.16415$&$1.35\rme-2$&$4.51\rme-6$&$1.46\rme-9$&$3.34\rme-2$\\[1.0ex]
$1.0$&$-2.875\rme-2$&$7.439\rme-2$&$1.77\rme-4$&$5.93\rme-7$&$1.35\rme-2$\\[1.0ex]
$0.1$&$-0.39508$&$0.23011$&$3.99\rme-3$&$6.39\rme-4$&$-0.1039$\\[1.0ex]
$0.01$&$-0.27014$&$0.16771$&$6.21\rme-3$&$-3.22\rme-3$&$-9.3\rme-2*$\\[1.0ex]
\hline\hline 
\end{tabular} 
\caption{\footnotesize Double derivatives of the 
contributions to O(3) wave functions for $\theta=0$ in all isospin 
channels. A double star ** indicates that there is no stability and a single
star * indicates a $\sim10\%$ variation between successive approximations.}  
\label{O3zeropot} 
\end{table}

\vfill
\eject


\eject

\end{document}